\documentclass[11pt]{article}
\usepackage{graphicx}
\usepackage{amssymb}
\def\lsim{\mathrel{\raise.3ex\hbox{$<$\kern-.75em\lower1ex\hbox{$\sim$}}}}
\def\gsim{\mathrel{\raise.3ex\hbox{$>$\kern-.75em\lower1ex\hbox{$\sim$}}}}

\newcommand{\be}{\begin{equation}}
\newcommand{\ee}{\end{equation}}
\newcommand{\bea}{\begin{eqnarray}}
\newcommand{\eea}{\end{eqnarray}}

\begin{document}

\thispagestyle{empty}

\title
{Lorentz noninvariant oscillations of massless neutrinos are excluded}
\author{V. Barger$^1$, Jiajun Liao$^2$, D. Marfatia$^{3,1}$ and K. Whisnant$^2$\\[2ex]
\small\it $^1$Department of Physics, University of Wisconsin, Madison, WI 53706, USA\\
\small\it $^2$Department of Physics and Astronomy, Iowa State University, Ames, IA 50011, USA \\
\small\it $^3$Department of Physics and Astronomy, University of Kansas, Lawrence, KS 66045, USA
}

\date{}

\maketitle

\begin{abstract}

The bicycle model of Lorentz noninvariant neutrino oscillations
without neutrino masses naturally predicts maximal mixing and a $1/E$
dependence of the oscillation argument for $\nu_\mu \to \nu_\tau$
oscillations of atmospheric and long-baseline neutrinos, but cannot
also simultaneously fit the data for solar neutrinos and KamLAND. Within 
the Standard Model Extension, we
examine all 19 possible structures of the effective Hamiltonian 
for Lorentz noninvariant oscillations of massless
neutrinos that naturally have a $1/E$ dependence at high neutrino
energy. Due to the lack of any evidence for direction dependence, we
consider only direction-independent oscillations. Although we find a
number of models with a $1/E$ dependence for atmospheric and
long-baseline neutrinos, none can also simultaneously fit solar and
KamLAND data.

\end{abstract}

\newpage

\section{Introduction}

Neutrino data from atmospheric, long-baseline, solar and reactor
experiments are easily explained by oscillations of three active,
massive neutrinos~\cite{review}. Lorentz-invariance and $CPT$
violating interactions originating at the Planck scale can also lead
to neutrino oscillations. The Standard Model Extension
(SME)~\cite{SME} includes all such interactions that may arise from
spontaneous symmetry breaking but still preserve Standard Model gauge
invariance and power-counting renormalizability. Studies of neutrino
oscillations with Lorentz invariance violation have been made both for
massive~\cite{Coleman, BPWW, tandem} and massless~\cite{K1, K2,
bmw-bike} neutrinos. A model with nonrenormalizable Lorentz invariance
violating interactions and neutrino mass has also been
proposed~\cite{Diaz}.  However, no viable model has been found that
does not require at least one nonzero neutrino mass. The purpose of
this paper is to determine if Lorentz invariance violation alone can
account for the verified oscillation phenomena seen in atmospheric,
long-baseline, solar and reactor neutrinos. We do not attempt to fit
the possible oscillation signals seen in the LSND~\cite{LSND} and
MiniBooNE~\cite{miniboone} experiments.

In the SME, the evolution of massless neutrinos in vacuum may be
described by the effective Hamiltonian~\cite{K1}
\be
(h_{eff})_{ij} = E \delta_{ij} + {1\over E} \left[ a_L^\mu p_\mu
- c_L^{\mu\nu} p_\mu p_\nu \right]_{ij} \,,
\label{eq:heff}
\ee
where $p_\mu = (E, -E\hat p)$ is the neutrino four-momentum, $\hat p$
is the neutrino direction, $i,j$ are flavor indices, and $a_L \to -a_L$
for antineutrinos. The coefficients $a_L$ have dimensions of energy
and the $c_L$ are dimensionless. Direction dependence of the neutrino
evolution enters via the space components of $a_L$ and $c_L$, $\mu$ or $\nu =
X, Y, Z$, while direction independent terms have $\mu = \nu = T$. The
Kronecker delta term on the right-hand side of Eq.~(\ref{eq:heff}) may
be ignored since oscillations are insensitive to terms in $h_{eff}$
proportional to the identity.

The two-parameter bicycle model~\cite{K1} can be defined as follows:
$(c_L)_{ij}$ has only one nonzero element in flavor space and the only
nonzero $(a_L)_{ij}$ are $(a_L)_{e\mu} = (a_L)_{e\tau}$. These
interactions can be nonisotropic, which could lead to different
oscillation parameters for neutrinos propagating in different
directions. In Ref.~\cite{bmw-bike} it was shown that the pure
direction-dependent bicycle model is ruled out by solar neutrino data
alone, while a combination of atmospheric, solar and long-baseline
neutrino data excludes the pure direction-independent case.  A mixture
of direction-dependent and direction-independent terms (with 5 parameters) 
is also excluded when KamLAND data are added~\cite{bmw-bike}.

The key feature of the bicycle model is that even though the terms in
$h_{eff}$ are either constant or proportional to neutrino energy, at
high neutrino energies there is a seesaw type mechanism that leads to
$1/E$ behavior for the oscillation argument for atmospheric and
long-baseline neutrinos. In this paper we examine the general case of
direction-independent Lorentz invariance violation in the Standard
Model Extension for three neutrinos without neutrino mass, {\it i.e.},
Eq.~(\ref{eq:heff}) with only $c_L^{TT}$ and $a_L^T$ terms. We do not
consider possible direction-dependent terms since there is no evidence
for direction dependence in neutrino oscillation experiments (see,
{\it e.g.}, the experiments in Ref.~\cite{auerbach} and the analysis of
Ref.~\cite{K1}). For notational simplicity we
henceforth drop the $L$ subscript and $T$ superscripts from the
$c_L^{TT}$ and $a_L^T$ in our formulae.

We first look for textures of the $c_{ij}$ in flavor space that allow
a $1/E$ dependence of the oscillation argument at high neutrino
energy.  We then check the phenomenology for atmopheric,
long-baseline, solar and reactor neutrino experiments. We were unable
to find any texture of $h_{eff}$ that could simultaneously fit all
the data.

In Sec.~2 we review the constraints on the direction-independent
bicycle model. In Sec.~3 we list all possible textures of the
$c$ coefficients and find which ones allow a $1/E$ dependence of the
oscillation argument at high neutrino energies. For those that do, we
first check the oscillation amplitude for atmospheric and long-baseline
neutrinos, and if suitable parameters are found we then check the
ability of the model to fit KamLAND and solar neutrino data. In Sec.~4
we summarize our results.

\section{Neutrino oscillations in the bicycle model}

As an illustrative analysis, we begin with a review of the
direction-independent bicycle model and show how it is inconsistent
with a combination of atmospheric, long-baseline and solar neutrino
data.

Neutrino oscillations occur due to eigenenergy differences in $h_{eff}$ and
the fact that the neutrino flavor eigenstates are not eigenstates of
$h_{eff}$. In our generalization of the direction-independent bicycle model,
\be
h_{eff} = \pmatrix{
-2cE + 2 a_{11} & a_{12} & a_{13} \cr
a_{12} & 0 & 0 \cr
a_{13} & 0 & 0} \,,
\ee
where the $c$ term is $CPT$-even and the $a_{ij}$ terms are $CPT$-odd.
The simple two-parameter bicycle model~\cite{K1} has $a_{13} = a_{12}$
and $a_{11} = 0$. We allow $a_{12}$ to be different from $a_{13}$ so
that mixing of atmospheric neutrinos may be (slightly) nonmaximal. The
$a_{11}$ term allows an adjustment of the oscillation probabilities of
low-energy solar neutrinos~\cite{K1}.

For this $h_{eff}$ there are two independent eigenenergy differences
$\Delta_{jk} = E_j - E_k$ given by
\be
\Delta_{21} =
\sqrt{(a_{11}-cE)^2 + a^2} + cE - a_{11} \,,\quad
\Delta_{32} =
\sqrt{(a_{11}-cE)^2 + a^2} - cE + a_{11} \,,
\label{eq:Delta}
\ee
where $a \equiv \sqrt{a_{12}^2 + a_{13}^2}$. The effective Hamiltonian is
diagonalized via $U^T h_{eff} U$ by the energy-dependent mixing matrix
\be
U = \pmatrix{-\cos\theta & 0 & \sin\theta \cr
\sin\phi\sin\theta & \cos\phi & \sin\phi\cos\theta \cr
\cos\phi\sin\theta & -\sin\phi &  \cos\phi\cos\theta} \,,
\label{eq:U0}
\ee
where
\bea
\sin^2\theta &=&
{1\over2}\left[ 1 + {a_{11} - cE\over\sqrt{(a_{11}-cE)^2 + a^2}} \right] \,,
\label{eq:theta}
\\
\tan\phi &=& {a_{12}\over a_{13}} \,.
\label{eq:phi}
\eea
The off-diagonal oscillation probabilities are
\bea
P(\nu_e \leftrightarrow \nu_\mu) &=&
\sin^2\phi \sin^22\theta \sin^2(\Delta_{31}L/2) \,,
\label{eq:Pem}
\\
P(\nu_e \leftrightarrow \nu_\tau) &=&
\cos^2\phi \sin^22\theta \sin^2(\Delta_{31}L/2) \,,
\label{eq:Pet}
\\
P(\nu_\mu \leftrightarrow \nu_\tau) &=&
\sin^2\theta \sin^22\phi \sin^2(\Delta_{21}L/2)
+ \cos^2\theta \sin^22\phi \sin^2(\Delta_{32}L/2)
\nonumber\\
&\phantom{=}& - {1\over4}\sin^22\phi \sin^22\theta \sin^2(\Delta_{31}L/2) \,,
\label{eq:Pmt}
\eea
where $\Delta_{31} = \Delta_{32} + \Delta_{21}$.

For large $E$, appropriate for atmospheric and long-baseline neutrinos, 
if $a^2 \ll (cE)^2$, then $\sin^2\theta \ll 1$, $\cos^2\theta \simeq 1$
and the only appreciable oscillation is
\be
P(\nu_\mu \leftrightarrow \nu_\tau) \simeq
\sin^22\phi \sin^2(\Delta_{32}L/2) \,,
\ee
where
\be
\Delta_{32} \simeq {a^2 \over 2cE} \,.
\label{eq:prob}
\ee
Thus the $\nu_\mu \to \nu_\tau$ oscillation amplitude has amplitude
$\sin^22\phi$ and is maximal for $\phi = {\pi\over 4}$, in which case
$a_{12} = a_{13}$ (reproducing the simple two-parameter bicycle model). The
energy dependence of the oscillation argument in this limit is the same as for
conventional neutrino oscillations due to neutrino mass differences, with an
effective mass-squared difference
\be
\delta m^2_{eff} = 2E \Delta_{32} = {a^2\over c} \,.
\label{eq:dm2}
\ee
The measured value for $\delta m^2_{eff}$ in atmospheric and
long-baseline experiments then places a constraint that relates $a$ and $c$.

If $E$ is not too large, then the more general Eqs.~(\ref{eq:U0})-(\ref{eq:phi})
apply. Furthermore, in matter there is an additional
term due to coherent forward scattering~\cite{matter}, which adds
a $\sqrt2 G_F N_e$ term to the upper left element of $h_{eff}$, where
$N_e$ is the electron number density. In matter the angle
$\phi$ is unchanged and $\theta$ is now given by Eq.~(\ref{eq:theta})
with the substitution $a_{11} \to a_{11} + G_F N_e/\sqrt2$. For
adiabatic propagation in the sun the solar neutrino oscillation probability is
\be
P(\nu_e \to \nu_e) = \cos^2\theta\cos^2\theta_0 + \sin^2\theta \sin^2\theta_0
\,,
\ee
where $\theta_0$ is the mixing angle at the creation point in the sun (with
electron number density $N_e^0 \simeq 90~N_A$/cm$^3$) and $\theta$ is
the mixing angle in vacuum.  For convenience we define the quantity
$b \equiv G_F N_e^0/(2\sqrt2) = 1.7\times10^{-12}$~eV.

The probability has a minimum value
\be
P_{min} = {1\over2} {a^2 \over a^2 + b^2} \,,
\ee
which is always less than ${1\over2}$. The minimum $P$ must match the
oscillation probability of the $^8$B neutrinos (which from the SNO
experiment~\cite{SNO} is $P_{min} \simeq 0.30$), which fixes $a$ to be
\be
a = b \sqrt{2P_{min} \over (1-2P_{min})} = 2.1\times10^{-12}{\rm~eV}\,.
\label{eq:a}
\ee

At very low energies the solar neutrino oscillation probability is
\be
P_{low} = {1\over2}\left[ 1 + {a_{11}(a_{11}+2b)
\over\sqrt{a_{11}^2 + a^2}\sqrt{(a_{11}+2b)^2+a^2}} \right] \,.
\label{eq:Plow}
\ee
Note that the probability in Eq.~(\ref{eq:Plow}) is exactly ${1\over2}$
for $a_{11} = 0$ ({\it e.g.}, in the simple two-parameter bicycle model),
which is not a good fit to the low-energy solar neutrino data.
However, for $a_{11} > 0$ or $a_{11} < - 2b$, the low-energy probability
can be made larger than ${1\over2}$. Using the low-energy value  $P \approx 1 -
{1\over2}\sin^22\theta_{12} \approx 0.57$, where $\theta_{12}$ is the usual
solar neutrino mixing angle~\cite{global-fits}, we find $a_{11} = 0.20b$
or $a_{11} = -2.2b$.

\begin{figure}[t]
\centering\leavevmode
\includegraphics[width=5in]{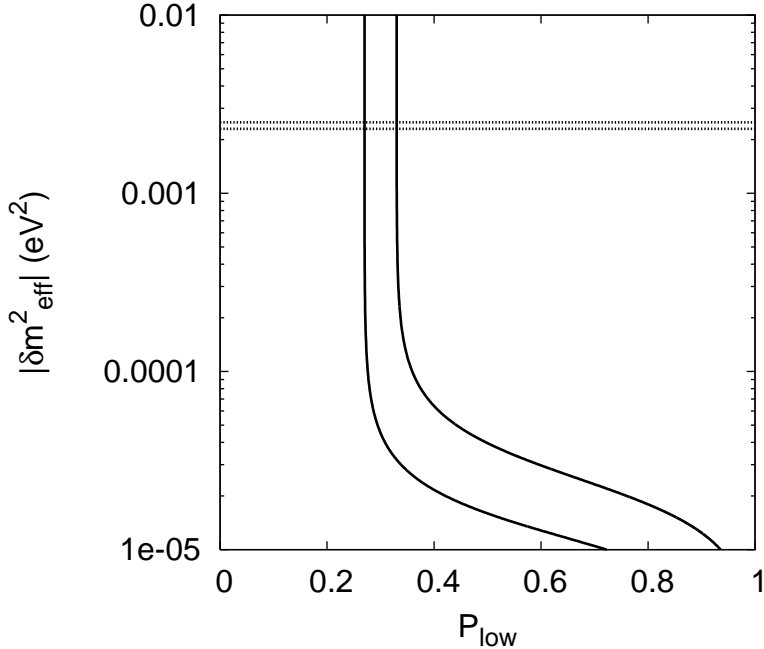}
\caption[]{Correlation of $\delta m^2_{eff}$ for high-energy atmospheric
and long-baseline neutrinos with the oscillation probability $P_{low}$
for low-energy solar neutrinos in the generalized direction-independent
bicycle model (between the solid curves), found by varying the model
parameter $a_{11}$. The left (right) solid curve assumes $E_{min} =
8.0$~MeV (12~MeV) and $P_{min} = 0.27~(0.33)$. The region between the
horizontal dotted lines is consistent with atmospheric and long-baseline neutrino
experiments.
\label{fig:correl}}
\end{figure}

The probability reaches the minimum at
\be
E_{min} = {1\over c}\left[ a_{11} + b \right] \,,
\ee
which must occur in the energy region of the $^8$B solar neutrinos
($E_{min} \approx 10$~MeV), which fixes the magnitude of $c$ to be
\be
|c| = {1\over E_{min}}\left| a_{11} + b \right| = {1.2~b\over E_{min}}
\approx 2.0\times10^{-19} \,.
\label{eq:c}
\ee
Using Eq.~(\ref{eq:dm2}) we may now calculate the value of the
atmospheric $\delta m^2_{eff}$ inferred from solar neutrino data:
$\delta m^2 = a^2/c = 2.2\times10^{-5}$~eV$^2$, which is two orders of
magnitude below the measured value.

One caveat for this calculation is that the low-energy solar oscillation
probability is not measured precisely, and the model prediction may be
adjusted by changing $a_{11}$. This in turn changes $c$ (via
Eq.~\ref{eq:c}) and the predicted atmospheric $\delta m^2_{eff}$ (via
Eq.~\ref{eq:dm2}).  The relationship between $\delta m^2_{eff}$ and
$P_{low}$ is shown in Fig.~\ref{fig:correl}, where we have assumed
8~MeV~$< E_{min} <$~12~MeV and 0.27~$< P_{min} <$~0.33. 

For the range of $\delta m^2_{eff}$ allowed by experiment (shown by
the horizontal dashed lines), the low-energy probability is
approximately $0.30$, which is not consistent with $P_{low} \approx
0.57$ preferred by the solar data. In fact, any $\delta m^2_{eff}$
above $10^{-4}$~eV$^2$ gives a value for $P_{low}$ below
0.40. Therefore there is no acceptable value for $a_{11}$ that fits
both the low-energy solar oscillation probability and $\delta
m^2_{eff}$ for high-energy atmospheric and long-baseline neutrinos,
and the generalized direction-independent bicycle model is excluded.

\section{Other textures for $h_{eff}$}

\subsection{Classification of models}

There are six possible $c$ coefficients in $h_{eff}$: three real
diagonal coefficients and three complex off-diagonal coefficients (the
remaining three off-diagonals are fixed by the hermiticity of
$h_{eff}$). Therefore there are $2^6 = 64$ possible $c$ textures for
$h_{eff}$. Since the high-energy behavior of $h_{eff}$ is determined
by the $c$ coefficients, we classify the models by the number of
nonzero $c$ there are in $h_{eff}$. Within each main class there are
distinct subclasses which depend on the diagonal/off-diagonal
structure; within each subclass there are textures that differ only by
permutation of the flavor indices. In all there are 19
subclasses, which are listed in Table~1.

\begin{table}[t]
\caption[]{A list of the 64 possible $c$ textures for $h_{eff}$.
The number in the subclass name corresponds to the number of nonzero
$c$, while the letters indicate a distinct diagonal/off-diagonal
structure (up to flavor permutation), if applicable. A $D_i$ in the
structure column indicates that a diagonal $c_{ii}$ is nonzero, while an
$O_{jk}$ indicates that off-diagonal $c_{jk}$ is nonzero. Different latin
indices in each case are distinct, {\it e.g.}, in the structure $D_i O_{jk}$
the diagonal element does not share a row or column with the
off-diagonal element, whereas for $D_{i} O_{ij}$ it does.}

\centering\leavevmode
\begin{tabular}{c l l c}\\
\hline
Number of & Subclass & Structure & Number of flavor\\
nonzero $c$ & & & permutations\\
\hline\hline
0 &  0 & $-$      & 1\\
\hline
1 & 1A & $D_i$    & 3\\
  & 1B & $O_{ij}$ & 3\\
\hline
2 & 2A & $D_i D_j$       & 3\\
  & 2B & $D_i O_{ij}$    & 6\\
  & 2C & $D_i O_{jk}$    & 3\\
  & 2D & $O_{ij} O_{ik}$ & 3\\
\hline
3 & 3A & $D_i D_j D_k$          & 1\\
  & 3B & $D_i D_j O_{ij}$       & 3\\
  & 3C & $D_i D_j O_{ik}$       & 6\\
  & 3D & $D_i O_{ij} O_{ik}$    & 3\\
  & 3E & $D_j O_{ij} O_{ik}$    & 6\\
  & 3F & $O_{ij} O_{ik} O_{jk}$ & 1\\
\hline
4 & 4A & $D_i D_j D_k O_{ij}$       & 3\\
  & 4B & $D_i D_j O_{ij} O_{ik}$    & 6\\
  & 4C & $D_i D_j O_{ik} O_{jk}$    & 3\\
  & 4D & $D_i O_{ij} O_{ik} O_{jk}$ & 3\\
\hline
5 & 5A & $D_i D_j D_k O_{ij} O_{ik}$    & 3\\
  & 5B & $D_i D_j O_{ij} O_{ik} O_{jk}$ & 3\\
\hline
6 &  6 & $D_i D_j D_k O_{ij} O_{ik} O_{jk}$ & 1\\
\hline\hline\\
\end{tabular}
\end{table}

We note that we may subtract any quantity proportional to the identity
from $h_{eff}$, since common phases in the neutrino equations of
motion do not affect the oscillations. In this way a diagonal element
may be removed or moved from one position to another. Then it is
not hard to see that the following subclasses are strictly equivalent:
3A$\leftrightarrow$2A, 3C$\leftrightarrow$3B, 4A$\leftrightarrow$3B,
4C$\leftrightarrow$4B, 5A$\leftrightarrow$4B and 6$\leftrightarrow$5B.

\subsection{Method for analyzing textures}

Our analysis proceeds as follows. We assume that $|c_{ij} E| \gg
|a_{k\ell}|$ for any $(i,j,k,\ell)$ for the high energies of
atmospheric and long-baseline neutrinos. This assumption is justified
since if any $a$ is similar in magnitude to the $cE$ at high
energies, then at lower energies (such as for reactor neutrinos) the
$a$ terms will dominate and the oscillation arguments will be
energy-independent, contrary to the KamLAND data, which measured a
spectral distortion (similarly, solar neutrinos would also not have an
energy-dependent oscillation probability, as they must).  Furthermore,
for the sake of naturalness, we assume that the $c$ coefficients are all the same order of
magnitude, and that likewise the $a$ coefficients are also the same
order of magnitude.

Although for each texture the number of nonzero $c$ is determined,
initially we place no restrictions on the $a$. We note that if all
off-diagonal $c$ are nonzero, then by a redefinition of neutrino
phases and adding a term proportional to the identity we may take all
off-diagonal $c$ to be real and positive, except for one off-diagonal
$c$ that is complex (which we take to be $c_{13}$ unless
otherwise noted). If any off-diagonal $c$ is zero, the nonzero
off-diagonal $c$ may all be taken as real and positive.

A key feature of the bicycle model was that even though the terms in
the effective Hamiltonian were either proportional to energy or
constant in energy, one eigenvalue difference was proportional to
$E^{-1}$, which mimics the energy dependence of the oscillations of
atmospheric and long-baseline neutrinos. Having an eigenvalue
difference proportional to $E^{-1}$ means that if the eigenvalues are
expanded in a power series in neutrino energy,
\bea
\lambda_i = \sum_{j=0}^\infty a_{ij} E^{1-j} \,,\qquad {\rm~for~}i=1,2,3 \,
\label{eq:lambda}
\eea
then two eigenvalues must be degenerate at leading order in $E$
(linear in $E$), and at the next order in energy ($E^0$, independent
of energy). Therefore in our analysis of more general three-neutrino
models with Lorentz invariance violation, we look for model parameters
that satisfy these conditions. Since an $L/E$ dependence has been seen
over many orders of magnitude in neutrino energy~\cite{superK}, it
seems likely that this is the only way the Hamiltonian in
Eq.~(\ref{eq:heff}) will be able to fit all atmospheric and
long-baseline neutrino data.

For each texture we expand the eigenvalues of $h_{eff}$
in powers of $E$ (as in Eq.~\ref{eq:lambda}),
where the leading $E^1$ behavior comes from the dominant $c E$
terms.  Since we want $1/E$ behavior for at least one oscillation
argument, we require that two of the eigenvalues be degenerate to
order $E^0$, with the first nonzero difference occurring at order
$E^{-1}$. In all cases this requirement puts constraints on the $c$
and $a$ coefficients. In our calculations we first find the
eigenvalues to order $E^1$ and impose the constraint that two
eigenvalues must be degenerate; then we find the eigenvalues of the
simplified $h_{eff}$ to order $E^0$ and again impose the degeneracy
condition. In this way the expressions for the eigenvalues to order
$E^{-1}$ will be made as simple as possible at each stage of the
calculation.

If the appropriate $1/E$ behavior can be achieved, the mixing angles
are then calculated to determine if $\nu_\mu$'s have maximal mixing
and $\nu_e$ small mixing for atmospheric and long-baseline
neutrinos. If the model is still viable, the energy dependences of
the oscillations of solar and KamLAND neutrinos are then 
checked for consistency.

At any time we are allowed to subtract a constant times the identity
matrix from $h_{eff}$. Some cases may then be further simplified, or
made equivalent to other cases (see below for specific examples).
Rotations are also sometimes used to show that some cases are
equivalent to others.

\subsection{No $c$ parameters}

In this case, Class 0, $h_{eff}$ has only $a$ terms and therefore is
independent of energy. This clearly cannot produce $1/E$ behavior at
high energy, so this category is immediately ruled out.

\subsection{One $c$ parameter}

\subsubsection{Class 1A}

This case has the structure
\bea
h_{eff} = \pmatrix{cE + a_{11} & a_{12} & a_{13} \cr
a_{12}^* & a_{22} & a_{23} \cr
a_{13}^* & a_{23}^* & a_{33}} \,,
\eea
where $c_{11} \equiv c$ may be taken as real and positive. The eigenvalues
to order $E^0$ are then
\bea
\lambda_1 = cE + a_{11} \,,\quad
\lambda_2,\lambda_3 = {1\over2}\left[a_{22} + a_{33}
\pm \sqrt{(a_{22}-a_{33})^2 + 4|a_{23}|^2} \right] \,.
\eea
The difference $\lambda_2 - \lambda_3$ can only be made zero to order $E^0$
if $a_{22} = a_{33}$ and $|a_{23}| = 0$. Then $a_{33}$ times the identity
may be subtracted from $h_{eff}$; if $a_{11}-a_{33}$ is redefined as
$a_{11}$, this case reduces to the generalized bicycle model described in
Sec.~2, which is excluded by the combined data.

\subsubsection{Class 1B}

This case has the structure
\bea
h_{eff} = \pmatrix{a_{11} & cE + a_{12} & a_{13} \cr
cE + a_{12}^* & a_{22} & a_{23} \cr
a_{13}^* & a_{23}^* & a_{33}} \,,
\eea
where $c_{12} \equiv c$ may be taken as real and positive. The eigenvalues
to order $E^1$ are then
\bea
\lambda_1, \lambda_2 = \pm|c|E \,,\quad
\lambda_3 = 0 \,.
\eea
Since these are all different at leading order, they cannot give an oscillation
argument proportional to $E^{-1}$ at high energies, and this case is not
allowed.

\subsection{Two $c$ parameters}

\subsubsection{Class 2A}

This case has the structure
\bea
h_{eff} = \pmatrix{c_{11}E + a_{11} & a_{12} & a_{13} \cr
a_{12}^* & c_{22}E + a_{22} & a_{23} \cr
a_{13}^* & a_{23}^* & a_{33}} \,,
\eea
where $c_{11}$ and $c_{22}$ are real. The eigenvalues at leading order
are
\bea
\lambda_1 = c_{11}E \,,\qquad
\lambda_2 = c_{22}E \,,\qquad
\lambda_3 = 0 \,,
\eea
so that we must have $c_{11} = c_{22}$ for degeneracy (having one of the
$c_{ii} = 0$ also works, but then it is in Class 1A instead of 2A). Now
if $c_{11}E$ times the identity is subtracted from $h_{eff}$, this reduces
to Class 1A, which is ruled out.

\subsubsection{Class 2B}

This case has the structure
\bea
h_{eff} = \pmatrix{c_{11}E + a_{11} & c_{12}E + a_{12} & a_{13} \cr
c_{12}E + a_{12}^* & a_{22} & a_{23} \cr
a_{13}^* & a_{23}^* & a_{33}} \,,
\eea
where $c_{11}$ and $c_{12}$ may be taken as real and positive. The
eigenvalues at leading order are
\bea
\lambda_1, \lambda_2 = {1\over2} \left[ c_{11} \pm
\sqrt{c_{11}^2 +4c_{12}^2} \right] E \,,\qquad \lambda_3 = 0 \,.
\eea
Degeneracy requires (i) $\lambda_1 = \lambda_2$, which is not possible
for nonzero $c_{11}$ and $c_{12}$, or (ii) $\lambda_3 = \lambda_1$ or
$\lambda_2$, which is not possible for nonzero $c_{12}$. Therefore
this case is not allowed.

\subsubsection{Class 2C}

This case has the structure
\bea
h_{eff} = \pmatrix{a_{11} & a_{12} & c_{13}E + a_{13} \cr
a_{12}^* & c_{22}E & a_{23} \cr
c_{13}E + a_{13}^* & a_{23}^* & a_{33}} \,,
\eea
where $c_{22}$ and $c_{13}$ may be taken as real and positive, and we have
subtracted a term proportional to the identity so that $a_{22} = 0$. The
eigenvalues at leading order are
\bea
\lambda_1, \lambda_2 = \mp c_{13} E \,,\qquad
\lambda_3 = c_{22} E \,.
\eea
Degeneracy requires $c_{13} = c_{22}$. If we define $c_{22} = c_{13}
\equiv  c$, where $c$ is a positive real number; then the eigenvalues
to order $E^0$ are
\bea
\lambda_1 =
-cE + {1\over2}\left[ a_{11} + a_{33} - 2 {\rm Re}(a_{13}) \right]
\,,\qquad
\lambda_2, \lambda_3 =
cE + {1\over4}\left[ x \pm \sqrt{x^2 + 8 |y|^2} \right] \,,
\eea
where $x \equiv a_{11} + a_{33} + 2{\rm Re}(a_{13})$
and $y \equiv a_{12} + a_{23}^*$. Degeneracy to order $E^0$ requires
$x = 0$ and $y = 0$, which implies $a_{11} + a_{33} = - 2{\rm Re}(a_{13})$
and $a_{12} = - a_{23}^*$. With these conditions the eigenvalues to
order $E^{-1}$ are
\bea
\lambda_1 &=& -cE + a_{11} + a_{33} - {1\over2cE}(2|a_{23}|^2 +
|a_{13}|^2 - a_{11} a_{33}) \,,
\\
\lambda_2 &=& cE + {1\over2cE}(2|a_{23}|^2 + |a_{13}|^2 - a_{11} a_{33}) \,,
\qquad
\lambda_3 = cE \,.
\eea
Clearly $\Delta_{32} = \lambda_3 - \lambda_2$ has the correct energy
dependence for atmospheric and long-baseline oscillations. The mixing
matrix such that $U^T h_{eff} U$ is diagonal at leading order is given by
\be
U = \pmatrix{
-{1\over\sqrt2} & {1\over\sqrt2}\sin\theta & {1\over\sqrt2}\cos\theta \cr
0 & \cos\theta & -\sin\theta \cr
{1\over\sqrt2} & {1\over\sqrt2}\sin\theta & {1\over\sqrt2}\cos\theta} \,,
\ee
where $\sin\theta = |a_{11} + a_{13}|/\sqrt{2|a_{23}|^2 + |a_{11} +
a_{13}|^2}$ and the oscillation probabilities are approximately given by
\bea
P(\nu_\mu \to \nu_\mu) &=&
1 - \sin^22\theta \sin^2\left({1\over2}\Delta_{32}L\right) \,,
\\
P(\nu_\mu \to \nu_e) &=& P(\nu_\mu \to \nu_\tau) =
{1\over2}\sin^22\theta \sin^2\left({1\over2}\Delta_{32}L\right) \,.
\eea
Therefore maximal mixing for $\nu_\mu$ is possible with $\delta m^2_{eff}
= 2E \Delta_{23} = (2|a_{23}|^2 + |a_{13}|^2 - a_{11}a_{33})/c$, but
$\nu_\mu$ oscillates equally to $\nu_e$ and $\nu_\tau$, which is
excluded by atmospheric neutrino experiments. Hence this case is not allowed.

\subsubsection{Class 2D}

This case has the structure
\bea
h_{eff} = \pmatrix{a_{11} & c_{12}E + a_{12} & c_{13}E + a_{13} \cr
c_{12}E + a_{12}^* & a_{22} & a_{23} \cr
c_{13}E + a_{13}^* & a_{23}^* & a_{33}} \,,
\eea
where $c_{12}$ and $c_{13}$ may be taken as real and positive. If a
rotation is applied to the $\mu-\tau$ sector, then $c_{13}$ may be rotated
away into $c_{12}$, which reduces this case to Class 1B, which is not allowed.

\subsection{Three $c$ parameters}

\subsubsection{Class 3A}

This subclass has nonzero $c$ in each diagonal term and no off-diagonal
$c$. By subtracting off $c_{33}E$ times the identity, this case reduces
to Class 2A, which is not allowed.

\subsubsection{Class 3B}

This case has the structure
\bea
h_{eff} = \pmatrix{c_{11}E + a_{11} & a_{12} & c_{13}E + a_{13} \cr
a_{12}^* & 0 & a_{23} \cr
c_{13}E + a_{13}^* & a_{23}^* & c_{33}E + a_{33}} \,,
\eea
where $c_{11}$, $c_{33}$ and $c_{13}$ may be taken as real
and $a_{22}$ has been set to zero by a subtraction proportional to the
identity. The eigenvalues at leading order are
\bea 
\lambda_1, \lambda_2 = {1\over2}\left[c_{11}+c_{33} \pm
\sqrt{(c_{11}-c_{33})^2+4c_{13}^2}\right] E \,, \qquad \lambda_3 = 0 \,.
\eea
There are two possible ways to have a degeneracy. First, if $\lambda_1 =
\lambda_2$, then we must have $c_{11} = c_{33}$ and $c_{13} = 0$. However,
if $c_{11}E$ times the identity is then subtracted from $h_{eff}$, this
possibility reduces to Class 1A. Second, we can have $\lambda_2 = 0$,
so that it is degenerate with $\lambda_3$. There is a family of such
solutions with $c_{33} = r^2 c_{11}$ and $c_{13} = r c_{11}$, where
$r$ may be taken as a positive real number. If we define
$c_{11} \equiv c$, then to order $E^0$ the eigenvalues are
\be
\lambda_1 = (1+r^2)cE + a_{11} + a_{33} - x \,,\qquad
\lambda_2, \lambda_3 = {1\over2}\left[x \pm \sqrt{x^2 + 4y} \right] \,,
\ee
where $x = [a_{33} + r^2 a_{11} - 2r{\rm Re}(a_{13})]/
(1+r^2)$ and $y = |r a_{12} - a_{23}^*|^2/(1+r^2)$. Degeneracy
is only possible if $x = 0$ and $y = 0$, which requires $a_{33} + r^2 a_{11}
= 2r{\rm Re}(a_{13})$ and $a_{23} = r a_{12}^*$, respectively. The
eigenvalues to order $E^{-1}$ are then
\bea
\lambda_1 &=& (1+r^2)cE + a_{11} + a_{33} +
{|a_{13}|^2 + (1+r^2) |a_{12}|^2 - a_{11}a_{33} \over(1+r^2)cE} \,,
\\
\lambda_2 &=& - {|a_{13}|^2 + (1+r^2) |a_{12}|^2 - a_{11}a_{33}
\over(1+r^2)cE} \,,
\qquad \lambda_3 = 0 \,.
\eea
Thus $\Delta_{32}$ has the correct energy dependence, and gives
\be
\delta m^2_{eff} = 2 E \Delta_{32} =
2 {|a_{13}|^2 + (1+r^2)|a_{12}|^2 - a_{11}a_{33} \over (1+r^2) c}
= 2 {|r a_{11} - a_{13}|^2 + (1+r^2)|a_{12}|^2\over c(1+r^2)} \,,
\label{eq:dm2eff}
\ee
for atmospheric and long-baseline neutrinos.  We note that
$\lambda_3 = 0$ is an exact result given the degeneracy conditions, true
even when $E$ is not large.

To leading order the mixing matrix that diagonalizes $h_{eff}$ via
$U^T h_{eff} U$ is
\bea
U = \pmatrix{\cos\phi & \sin\phi\cos\theta & -\sin\phi\sin\theta \cr
0 & \sin\theta & \cos\theta \cr
\sin\phi & -\cos\phi\cos\theta & \cos\phi\sin\theta} \,,
\label{eq:U3B}
\eea
where $\sin\phi \equiv r/\sqrt{1+r^2}$ and $\tan\theta \equiv \sqrt{1+r^2}
|a_{12}|/|r a_{11} - a_{13}|$. This mixing gives the
oscillation probabilities
\bea
P(\nu_\mu \to \nu_\mu) &=&
1 - \sin^22\theta \sin^2\left(\Delta_{32}{L\over2} \right) \,,
\\
P(\nu_\mu \to \nu_e) &=&
\sin^2\phi \sin^22\theta \sin^2\left(\Delta_{32}{L\over2} \right) \,,
\\
P(\nu_e \to \nu_e) &=&
1 - \cos^2\theta\sin^22\phi \sin^2 \left(\Delta_{21}{L\over2} \right)
  - \sin^2\theta\sin^22\phi \sin^2 \left(\Delta_{31}{L\over2} \right)
\nonumber
\\
&\phantom{=}& \phantom{1 -}
-\sin^4\phi\sin^22\theta \sin^2 \left(\Delta_{32}{L\over2} \right) \,.
\eea
Maximal $\nu_\mu$ oscillations are possible for $\theta \simeq \pi/4$,
which imposes the condition $\sqrt{1+r^2}|a_{12}| \simeq |ra_{11} - a_{13}|$.

Oscillations of $\nu_e$ at high energies must be small due to the
limit on $\nu_\mu \to \nu_e$ from K2K~\cite{K2K-nue} and
MINOS~\cite{MINOS-nue}.\footnote{Limits on $\nu_\mu \to \nu_e$ or
$\bar\nu_\mu \to \bar\nu_e$ from experiments such as CHOOZ or KARMEN
do not apply here since they involve lower energy neutrinos. MiniBooNE
limits may apply, but only for $\delta m^2_{eff} \gsim 0.1$~eV$^2$,
and therefore not at the $\Delta_{32}$ scale.} For K2K and MINOS the
oscillation amplitude for $P(\nu_\mu \to \nu_e)$, $\sin^2\phi
\sin^22\theta$, has an upper bound of about 0.14, which implies $r <
0.43$ for $\theta \simeq \pi/4$. The T2K experiment sees evidence for
$\nu_\mu \to \nu_e$ at the $2.5\sigma$ level~\cite{T2K}; their
allowed regions are consistent with this bound.

We note that the conditions $c_{33} = r^2 c_{11}$ and $c_{13} = r
c_{11}$ require fine tuning. If these conditions are not exact, they
introduce small corrections, which may be absorbed into the $a$
terms, {\it e.g.}, $a_{ij} \to a_{ij} + \delta c_{ij}E$, where $\delta
c_{ij}$ represents the deviation from the exact degeneracy
condition. This effectively introduces an $E$ dependence into $\delta
m^2_{eff}$, contrary to the atmospheric and long-baseline data.


For solar or reactor neutrinos the large energy limit does not apply.
Then the eigenvalues are
\bea
\lambda_1, \lambda_2 &=& {1\over2} \left[
cE(1+r^2) + a_{11}+ a_{33} \pm
\sqrt{[cE(1+r^2)+ a_{11} + a_{33}]^2 + 2\delta m^2_{eff}c(1+r^2)} \right]
\nonumber
\\
\qquad \lambda_3 &=& 0 \,.
\eea
where $\delta m^2_{eff}$ is from Eq.~(\ref{eq:dm2eff}), and it can be shown
that the matrix that diagonalizes $h_{eff}$ is
\be
U = \pmatrix{
\cos\phi\cos\xi+\sin\phi\cos\theta\sin\xi e^{-i\delta}
& -\cos\phi\sin\xi+\sin\phi\cos\theta\cos\xi e^{-i\delta}
& - \sin\phi\sin\theta
\cr
\sin\theta\sin\xi & \sin\theta\cos\xi & \cos\theta e^{i\delta}
\cr
\sin\phi\cos\xi-\cos\phi\cos\theta\sin\xi e^{-i\delta}
& -\sin\phi\sin\xi-\cos\phi\cos\theta\cos\xi e^{-i\delta}
& \cos\phi\sin\theta} \,,
\label{eq:U3B2}
\ee
where $\phi$ and $\theta$ are defined as above, $\tan\xi =
|ra_{11}-a_{13}|/(\lambda_1\cos\theta)$ and $\delta =
{\rm~arg} (r a_{11} - a_{13})$. Note that in the large energy limit
$\lambda_1$ is large, $\xi \to 0$, and Eq.~(\ref{eq:U3B2}) reduces to
Eq.~(\ref{eq:U3B}). Also, since none of the mixings are
zero, $CP$ violation is possible.

\begin{figure}
\centering\leavevmode
\includegraphics[width=10cm]{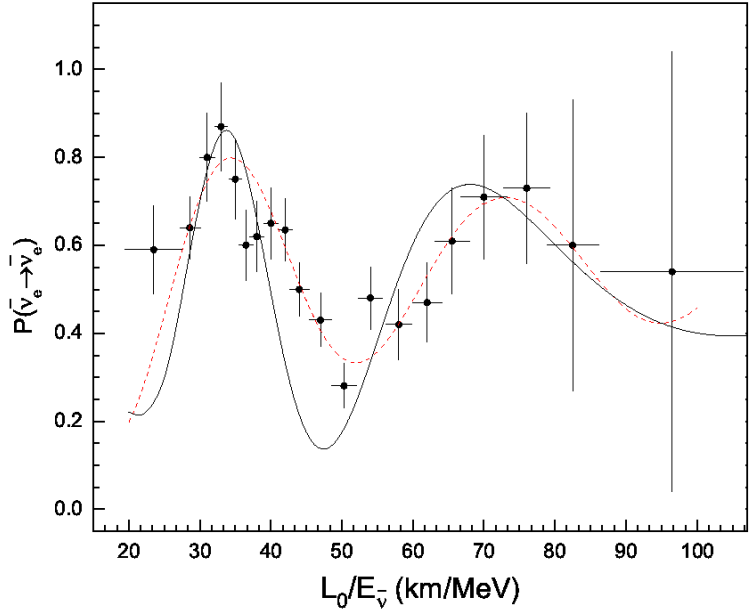}
\caption{The best fit to the KamLAND data for Class 3B (solid lines) and
the standard oscillation scenario with neutrino masses (dashed lines).
For the model parameters see Eq.~(\ref{modp}).}
\label{fig:3BKamLAND}
\end{figure}

We checked KamLAND phenomenology first.  Since we have obtained
several conditions from fitting the atmospheric and long-baseline
neutrinos, using these conditions we can vary \(a_{11}\),
\(a_{13}\) and \(r\) to fit the KamLAND data~\cite{kamland}. Other
parameters in the effective Hamiltonian will be determined by these three
parameters. Scanning the \(a_{11}\), \(a_{13}\) and \(r\)
parameter space, we find the following parameter values yield
reasonable agreement with the KamLAND data (see
Fig.~\ref{fig:3BKamLAND}):
\be
a_{11} = -8.7\times10^{-11}{\rm~eV} \,,\ \ \ \ \ 
a_{13} = -9.5\times10^{-11}{\rm~eV} \,, \ \ \ \ \ 
r = 0.1 \,.
\label{modp}
\ee
However, the fit is not as good as the standard oscillation scenario
with neutrino mass.

\begin{figure}[!htb]
\centering\leavevmode
\includegraphics[width=10cm]{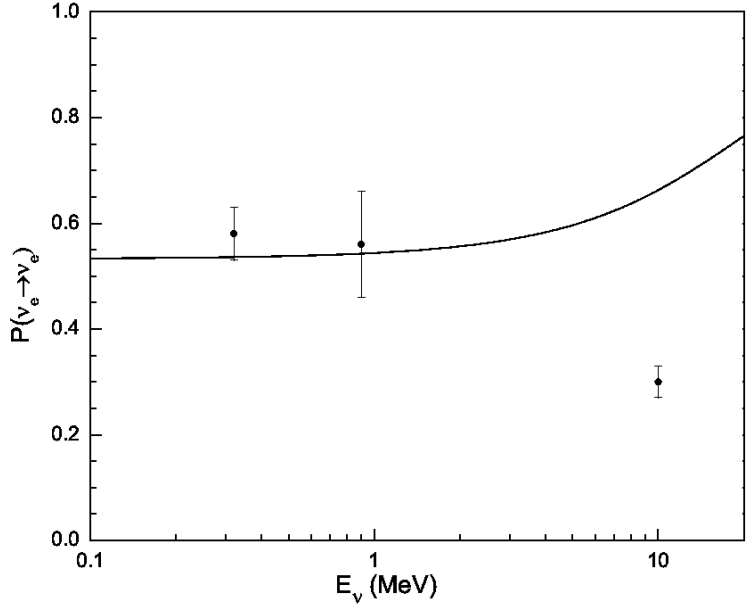}
\caption{The prediction of Class 3B for the oscillation probability
of solar neutrinos using the parameter values obtained from fitting
KamLAND data~\cite{kamland}. The solar data points are from
an update of the analysis in Ref.~\cite{BMW}.}
\label{fig:3Bsolar1}
\end{figure}

Next we use these parameter values to check the solar
phenomenology. Since the operator for \(a\) breaks $CPT$, we reverse
the sign of \(a\) when we apply these parameter values to the solar
neutrinos. However, the prediction does not agree with the solar data
at high energies given the upper bound on $r$ from above (see
Fig.~\ref{fig:3Bsolar1}).

\begin{figure}[!htb]
\centering\leavevmode
\includegraphics[width=10cm]{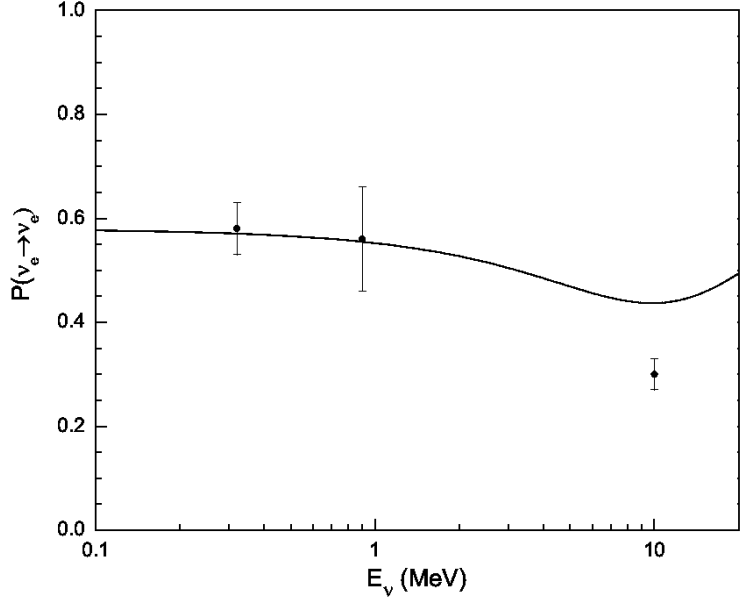}
\caption{Best fit prediction for survival probability of \(\nu_{e}\)
for solar neutrinos for Class 3B, assuming $r < 0.43$. The model
parameters for the best fit are \(a_{11}=-1.0\times10^{-10}{\rm~eV},
a_{13}=7.3\times10^{-11}{\rm~eV}\) and \(r=0.4\).}
\label{fig:3Bsolar2}
\end{figure}

\begin{figure}[!htb]
\small
\centering
\includegraphics[width=10cm]{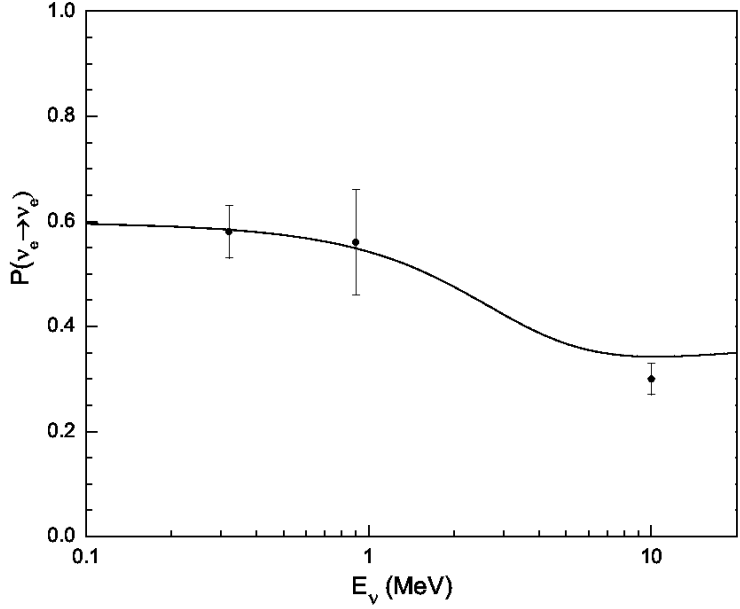}
\caption{Best fit prediction for survival probability of \(\nu_{e}\)
for solar neutrinos alone in Class 3B, assuming \(r > 0.43\). The model
parameters are \(a_{11} = -5.0\times10^{-10}{\rm~eV}, a_{13} =
-3.1\times10^{-11}\)~eV and \(r=1.0\).}
\label{fig:3Bsolar3}
\end{figure}

We also searched the \(a_{11}\), \(a_{13}\) and \(r\) parameter
space to fit the solar data separately. The best fit still can not
yield reasonable agreement with the solar data at high energies for
$r < 0.43$ (see Fig.~\ref{fig:3Bsolar2}).\footnote{In order to understand why the oscillation probability for the
high-energy solar neutrinos is so high, we consider the survival
probability of solar neutrinos in the high energy limit. As we have
noted, the mixing matrix in vacuum reduces to Eq.~(\ref{eq:U3B}) in the high
energy limit. In matter, we can still write the mixing matrix in the form,
\be
U_0 =
\pmatrix{
\cos\phi &  \sin\phi \cos\theta_0 & - \sin\phi \sin\theta_0
\cr
0 & \sin\theta_0  & \cos\theta_0
\cr
\sin\phi & - \cos\phi \cos\theta_0 &  \cos\phi \sin\theta_0
} \,,
\ee
except \(\tan\theta_0 \equiv r \sqrt{1+r^2}|a_{12}|/|a_{33}-r a_{13}^*|\),
since we do not have the relation \(a_{33} +r^2 a_{11} = 2r {\rm Re}(a_{13})\)
in matter (but $\theta_0$ is still the same as $\theta$ in vacuum). Now the
survival probability of the solar neutrinos in the high energy limit is
\be
P(\nu_e \rightarrow \nu_e) =
\cos^4\phi + \frac{1}{2} \sin^4\phi (1 + \cos2\theta \cos2\theta_0) \,.
\ee
Since \(\theta_0 = \theta \simeq \pi /4\) in vacuum, we have
\be
P(\nu_e \rightarrow \nu_e) = \cos^4\phi + \frac{1}{2} \sin^4\phi =
\frac{3}{2}(\sin^2\phi- \frac{2}{3})^2+\frac{1}{3} \,.
\ee
Since \(\sin \phi \equiv r/ \sqrt{1+r^2}\), applying the constraint
for \(r\), \(r < 0.43\) gives \(\sin^2 \phi < 0.14\), and the
$\nu_e$ survival probability approaches 0.75 in the high energy
limit. This is the reason that we cannot fit the solar data at 
high energies with the constraint on \(r\).} If we do not impose the
constraint on \(r\), the fit to the solar data is improved at high
energies (see Fig.~\ref{fig:3Bsolar3}). However, we cannot 
simultaneously fit the KamLAND and solar data even with larger
\(r\). We found that we also need \(|a_{11}|\) to become larger in
order to fit the solar data, but larger \(|a_{11}|\) yields fast
oscillations for KamLAND data with averaged probabilities around
\(1/2\).

\subsubsection{Class 3C}

This case has the structure
\bea
h_{eff} = \pmatrix{c_{11}E + a_{11} & a_{12} & c_{13}E + a_{13} \cr
a_{12}^* & c_{22}E + a_{22} & a_{23} \cr
c_{13}E + a_{13}^* & a_{23}^* & a_{33}} \,,
\eea
where $c_{11}$ and $c_{22}$ are real and $c_{13}$ may be taken as real
and positive. By subtracting $c_{22}$ times the identity, this case
reduces to Class 3B, which was described in the previous section.

\subsubsection{Class 3D}

This case has the structure
\bea
h_{eff} = \pmatrix{c_{11}E + a_{11} & c_{12}E + a_{12} & c_{13}E + a_{13} \cr
c_{12}E + a_{12}^* & a_{22} & a_{23} \cr
c_{13}E + a_{13}^* & a_{23}^* & a_{33}} \,,
\eea
where $c_{11}$, $c_{12}$ and $c_{13}$ may be taken as real and
positive.  If a rotation is applied to the $\mu-\tau$ sector, then $c_{13}$
may be rotated away, which reduces this case to Class 2B, which is not
allowed.

\subsubsection{Class 3E}

This case has the structure
\bea
h_{eff} = \pmatrix{c_{11}E + a_{11} & c_{12}E + a_{12} & a_{13} \cr
c_{12}E + a_{12}^* & a_{22} & c_{23}E + a_{23} \cr
a_{13}^* & c_{23}E + a_{23}^* & a_{33}} \,,
\eea
where $c_{11}$, $c_{12}$ and $c_{23}$ may be taken as real and positive.
This is the first case that cannot be simply reduced to a previous case,
and which requires solving a nontrivial cubic equation to determine the
eigenvalues at leading order. The eigenvalue equation for $h_{eff}/E$ at
leading order is
\bea
\lambda^3 - c_{11} \lambda^2 - (c_{12}^2 + c_{23}^2) \lambda
+ c_{11} c_{23}^2 = 0 \,.
\eea

For a cubic equation of the form $\lambda^3 + a\lambda^2 + b\lambda +
c = 0$, if we define $q=a^2-3b$ and $r = 2a^3-9ab+27c$, then
for three real roots the cubic discriminant $f \equiv 4q^3 - r^2$ must be
nonnegative, with $f=0$ when two of the roots are equal. Since the
effective Hamiltonian is hermitian, the eigenvalues must be real, so $f \ge
0$. Therefore, if there is a degeneracy, not only must $f =
0$, it must be a global minimum of $f$, {\it i.e.}, we can search for
degeneracies by finding the minima of $f$.

For this case we have
\be
q = c_{11}^2 + 3(c_{12}^2 + c_{23}^2) \,,\qquad
r = c_{11} (-2 c_{11}^2 +18 c_{23}^2 - 9 c_{12}^2)
\,,
\ee
and the discriminant is
\be
f = 4 (c_{11}^2 + 3 c_{12}+ 3 c_{23}^2)^3
- c_{11}^2 (2 c_{11}^2 + 9 c_{12}^2 - 18 c_{23}^2)^2
\,.
\ee
Then,
\be
0 = {\partial f\over\partial c_{12}} =
108 c_{12} \left[ 6 c_{12}^4 + 6 c_{23}^4 + c_{11}^2 c_{12}^2
+ 10 c_{11}^2 + 12 c_{12}^2 c_{23}^2 \right] \,,
\ee
requires at least that $c_{12} = 0$, which reduces this case to
Class 2C, which is not allowed.

\subsubsection{Class 3F}

This case has the structure
\bea
h_{eff} = \pmatrix{a_{11} & c_{12}E + a_{12} & c_{13}E + a_{13} \cr
c_{12}E + a_{12}^* & a_{22} & c_{23}E + a_{23} \cr
c_{13}^*E + a_{13}^* & c_{23}E + a_{23}^* & 0} \,,
\eea
where $c_{12}$ and $c_{23}$ may be taken as real and positive, $c_{13}$
is complex and $a_{33}$ has been set equal to zero. This case also
requires solving a nontrivial cubic equation to find the eigenvalues
at leading order; with
\be
q = 3(c_{12}^2 + c_{13}^2 + c_{23}^2) \,,\qquad
r = -54 c_{12} c_{23} |c_{13}| c_\delta \,,
\ee
where $c_\delta =\cos\delta$ and $\delta$ is the phase of $c_{13}$.
Searching for a minimum of $f = 4q^3-r^2$:
\bea
0 = {\partial f \over \partial c_{12}} &=&
72 c_{12} q^2 + 108 r |c_{13}| c_{23} c_{\delta} \,,
\label{eqn:dfdc12}
\\
0 = {\partial f \over \partial c_{13}} &=&
72 |c_{13}| q^2 + 108 r c_{12} c_{23} c_{\delta} \,,
\label{eqn:dfdc13}
\\
0 = {\partial f \over \partial c_{23}} &=&
72 c_{23} q^2 + 108 r c_{12} |c_{13}| c_{\delta} \,,
\label{eqn:dfdc23}
\\
0 = {\partial f \over \partial \delta} &=&
- 108 r c_{12} |c_{13}| c_{23} \sin\delta \,.
\eea
The quantity $q$ is explicitly nonzero; if $r$ was zero then
Eqs.~(\ref{eqn:dfdc12})-(\ref{eqn:dfdc23}) would imply
that $c_{12}$, $c_{13}$ and $c_{23}$ would all have to be zero, which is not
possible for this case, so $r \ne 0$. Then the last equation implies
$\sin\delta = 0$, or $\delta = 0$ or $\pi$. Thus $c_\delta = \pm 1$,
{\it i.e.}, $c_{13}$ is real, but it might differ by a sign from $c_{12}$ and
$c_{23}$; we use $c_{\delta}$ to denote this possible sign difference and
henceforth take $c_{13}$ as real and positive.

It is not hard to show that $c_{12} = c_{13} = c_{23}$ is required
for a minimum of $f$ and that this condition gives $f=0$. Therefore degeneracy
requires $c_{12} = c_{13} = c_{23} \equiv c$ with $c_\delta = \pm 1$.
Then the eigenvalues to order $E^0$ are
\be
\lambda_1 = 2cEc_\delta + a_{11} + a_{22} - x\,, \qquad
\lambda_2, \lambda_3 =
-cEc_\delta + {1\over2}\left[ x \mp \sqrt{x^2 - 4y} \right] \,,
\ee
where
\bea
x &=& {2\over3} \left[(a_{11} + a_{22}) - {\rm Re}(a_{13})
- c_\delta{\rm Re}(a_{12} + a_{23}) \right],,
\\
y &=& {1\over3}\left[ a_{11}a_{22} - |a_{12}|^2- |a_{13}|^2- |a_{23}|^2
+ 2{\rm Re}(a_{12}a_{23}-a_{22}a_{13}) \right.
\nonumber
\\
&\phantom{=}& \phantom{~{1\over3} [} +
\left. 2 c_\delta {\rm Re}( a_{12}a_{13}^*
+ a_{23} a_{13}^* - a_{11}a_{23}) \right] \,.
\eea
Thus degeneracy requires that the quadratic discriminant $g = x^2 - 4y$ be
zero. Since the eigenvalues are real, we know $g \ge 0$, and degeneracy
can only occur at a minimum of $g$. It can be
shown that $g$ has a minimum at zero for $a_{11} = a_{22} = {\rm Re}(a_{12})
= {\rm Re}(a_{13}) = {\rm Re}(a_{23}) = 0$ and ${\rm Im}(a_{13}) =
c_\delta {\rm Im}(a_{12} + a_{23})$. Then
\be
h_{eff} = \pmatrix{0 & cE +ia_{12} & c_\delta[cE+i(a_{12}+a_{23})] \cr
cE - ia_{12} & 0 & cE + ia_{23} \cr
c_\delta[cE - i(a_{12}+a_{23})] & cE - ia_{23} & 0} \,,
\ee
where the $a_{ij}$ are now defined as real. The eigenvalues of this matrix
to order $E^{-1}$ are
\be
\lambda_1 = 2c_\delta cE + {2c_\delta\over3cE}(a_{12}^2 + a_{23}^2
+ a_{12} a_{23}) \,,\qquad
\lambda_2 = - c_\delta cE \,, \qquad
\lambda_3 = - c_\delta cE - {2c_\delta\over3cE}(a_{12}^2 + a_{23}^2
+ a_{12} a_{23}) \,,
\ee
and the mixing matrix that diagonalizes $h_{eff}$ is
\be
U = \pmatrix{{1\over\sqrt3} & {1\over N} a_{23}
& - {c_\delta\over\sqrt3 N}(a_{23} + 2a_{12}) \cr
{c_\delta\over\sqrt3} & - {c_\delta\over N} (a_{12} + a_{23})
& {1\over\sqrt3 N}(a_{12} - a_{23}) \cr
{1\over\sqrt3} & {1\over N} a_{12}
& {c_\delta\over\sqrt3 N}(a_{12} + 2a_{23})} \,,
\label{eq:U3F}
\ee
where $N = \sqrt{2(a_{12}^2 + a_{23}^2 + a_{12}a_{23})}$ is a normalization
factor. 

At high energies for the fast oscillation \(\Delta_{31} \simeq
\Delta_{21} \simeq -3 c_\delta c E\), all off-diagonal
oscillation probabilities have the same approximate form:
\begin{equation}
P(\nu_\alpha \rightarrow \nu_\beta) =
\frac{4}{9}\sin^2\left(\frac{3 c E L}{2} \right) \,.
\end{equation}
For this oscillation amplitude, 4/9, the NuTeV limit on $\nu_\mu \to
\nu_e$~\cite{nutev} gives a 90\%~C.L. upper bound on \(\delta
m_{eff}^2\) of 3.6~eV$^2$. Since we have \(\delta m_{eff}^2=6 c E^2\)
in this case and the average neutrino energy was 74~GeV, the
experiment imposes the upper bound \(c \leq 1.1 \times 10^{-22}
\).\footnote{The NuTeV bound on $\delta m^2$ is not the most stringent
for ordinary massive neutrino oscillations, but because $\delta
m^2_{eff} \propto E^2$, the high neutrino energies in NuTeV give the
strongest bound on $c$.} On the other hand, in order for the expansion
in powers of $E$ to be valid, we need $N/(cE) \ll 1$ for $E \gsim
100$~MeV, which leads to the lower bound $c > 4\times10^{-17}$.
Therefore the structure required for the $1/E$ behavior at high energy
is inconsistent with accelerator bounds.  Since all flavors have the
same survival probability in the fast oscillation, the result is the
same even if a different permutation of flavors is taken.

Furthermore, for the \(\Delta_{23}\) oscillations in atmospheric and
long-baseline neutrinos, all three flavors have the probability
\begin{equation}
P(\nu_\alpha \rightarrow \nu_\alpha) = 
\frac{5}{9}-4 {|U_{\alpha 2}|}^2 \left(\frac{2}{3}-{|U_{\alpha 2}|}^2 \right)
\sin^2\left(\frac{\Delta_{23} L}{2} \right) \,.
\end{equation}
This implies that all flavors of downward atmospheric neutrinos would
be suppressed by a factor of 5/9, which is contrary to the data.
Therefore this case is excluded.


\subsection{Four $c$ parameters}

\subsubsection{Class 4A}

This case has three nonzero diagonal and one nonzero diagonal $c$. By
subtracting a piece proportional to the identity, this case may be reduced
to either Class 3B or 3C.

\subsubsection{Class 4B}

This case has the structure
\bea
h_{eff} = \pmatrix{c_{11}E + a_{11} & c_{12}E + a_{12} & c_{13}E + a_{13} \cr
c_{12}E + a_{12}^* & c_{22}E + a_{22} & a_{23} \cr
c_{13}E + a_{13}^* & a_{23}^* & a_{33}} \,,
\eea
where $c_{11}$ and $c_{22}$ are real and $c_{12}$ and $c_{13}$ may be taken as
positive. The eigenvalue equation for $h_{eff}/E$ at leading order is
\bea
\lambda^3 - (c_{11} + c_{22}) \lambda^2 + (c_{11} c_{22} - c_{12}^2 - c_{23}^2)
\lambda + c_{22} c_{13}^2 = 0 \,.
\eea
This case has cubic discriminant
\be
f = 4 q^3 - r^2 \,,
\ee
where
\bea
q &\equiv& c_{11}^2 + c_{22}^2 -c_{11} c_{22} + 3 c_{12}^2 + 3 c_{13}^2 \,,
\\
r &\equiv& -2c_{11}^3 + 3 c_{11}^2 c_{22} + 3 c_{11} c_{22}^2 - 2 c_{22}^3
- 9 c_{11} (c_{12}^2 + c_{13}^2) + 9 c_{22} (2 c_{13}^2 - c_{12}^2) \,,
\eea
The minimum conditions are
\bea
0 &=& {\partial f\over\partial c_{11}} =
12(2c_{11} - c_{22})q^2 - 2r \left[-6c_{11}^2 + 6 c_{11}c_{22} + 3 c_{22}^2
-9(c_{12}^2 + c_{13}^2) \right] \,,
\label{eq:1a}\\
0 &=& {\partial f\over\partial c_{22}} =
12(2c_{22} - c_{11})q^2 - 2r \left[-6c_{22}^2 + 6 c_{22}c_{11} + 3 c_{11}^2
+9(2c_{13}^2 - c_{12}^2) \right] \,,
\label{eq:2a}\\
0 &=& {\partial f\over\partial c_{12}} =
72 c_{12} q^2 + 36r c_{12} (c_{11} + c_{22}) \,,
\label{eq:3a}\\
0 &=& {\partial f\over\partial c_{13}} =
72 c_{13} q^2 - 36r c_{13} (2c_{22} - c_{11}) \,.
\label{eq:4a}
\eea
Clearly $q > 0$ if none of the $c_{ij}$ are zero. If $r=0$, then
Eqs.~(\ref{eq:3a}) and (\ref{eq:4a}) would imply $c_{12} = c_{13} = 0$,
which is not Class 4B; therefore $r\ne0$. Then Eqs.~(\ref{eq:3a}) and
(\ref{eq:4a}) imply
\be
-{1\over c_{11} +c_{22}} = {r\over q^2} =
{1\over 2c_{22} - c_{11}} \,,
\ee
which implies $c_{22} = 0$. This case then reduces to Class 3D, which is not
allowed.

\subsubsection{Class 4C}

This case has the structure
\bea
h_{eff} = \pmatrix{c_{11}E + a_{11} & a_{12} & c_{13}E + a_{13} \cr
a_{12}^* & c_{22}E + a_{22} & c_{23}E + a_{23} \cr
c_{13}E + a_{13}^* & c_{23}E + a_{23}^* & a_{33}} \,,
\eea
where $c_{11}$ and $c_{22}$ are real and $c_{13}$ and $c_{23}$ may be taken
as real and positive. By subtracting $c_{11}$ times the identity this case may
be reduced to Class 4B, which is not allowed.

\subsubsection{Class 4D}

This case has the structure
\bea
h_{eff} = \pmatrix{c_{11}E + a_{11} & c_{12}E + a_{12} & c_{13}E + a_{13} \cr
c_{12}E + a_{12}^* & a_{22} & c_{23}E + a_{23} \cr
c_{13}^*E + a_{13}^* & c_{23}E + a_{23}^* & a_{33}} \,,
\eea
where $c_{11}$, $c_{12}$ and $c_{23}$ may be taken as real, and $c_{13}$
is complex. The eigenvalue equation for $h_{eff}/E$ at leading order is
\be
\lambda^3 - c_{11} \lambda^2 + - (c_{12}^2 + c_{13}^2 + c_{23}^2)
\lambda + c_{11} c_{23}^2 - 2 c_{12} c_{13} c_{23} c_\delta = 0 \,,
\ee
where $c_\delta \equiv \cos\delta$, $c_{13} \to c_{13}e^{i\delta}$ and
$c_{13}$ is now taken as real and positive. This case has
\bea
q &\equiv& c_{11}^2 + 3 (c_{12}^2 + c_{13}^2 + c_{23}^2) \,,
\\
r &\equiv& -2c_{11}^3 + 9 c_{11} (2 c_{23}^2 -c_{12}^2 - c_{13}^2)
- 54 c_{12}c_{13}c_{23}c_\delta \,,
\eea
where the discriminant is $f = 4 q^3 - r^2$. The minimum conditions are
\bea
0 &=& {\partial f\over\partial c_{11}} =
24 c_{11} q^2 - 2r \left[-6c_{11}^2  + 9(2 c_{23}^2 - c_{12}^2 - c_{13}^2)
\right] \,,
\label{eq:1b}\\
0 &=& {\partial f\over\partial c_{12}} =
72 c_{12} q^2 + 36r (c_{11}c_{12} + 3 c_{13}c_{23}c_\delta) \,,
\label{eq:2b}\\
0 &=& {\partial f\over\partial c_{13}} =
72 c_{13} q^2 + 36r (c_{11}c_{13} + 3 c_{12}c_{23}c_\delta) \,,
\label{eq:3b}\\
0 &=& {\partial f\over\partial c_{23}} =
72 c_{23} q^2 - 36r (2c_{11}c_{23} - 3 c_{12}c_{13}c_\delta) \,,
\label{eq:4b}\\
0 &=& {\partial f\over\partial\delta} =
-108 r c_{12}c_{13}c_{23}\sin\delta \,.
\label{eq:5b}
\eea
Clearly $q > 0$ if none of the $c$ are zero. If $r=0$, then
Eqs.~(\ref{eq:2b})-(\ref{eq:4b}) would imply $c_{12} = c_{13} = c_{23}
= 0$, which is not Class 4D; therefore $r\ne0$. Thus Eq.~(\ref{eq:5b})
implies $\sin\delta = 0$, or $c_\delta = \pm 1$; therefore the
off-diagonal elements are real.

By combining Eqs.~(\ref{eq:2b}) and (\ref{eq:3b}), we find $c_{12} =
c_{13}$, and by combining Eqs.~(\ref{eq:2b}) and (\ref{eq:4b}), we find
$c_{12}^2 = c_{23}^2 + c_{11}c_{23}c_\delta$. Then $q = (c_{11} + 3 c_\delta
c_{23})^2$ and $r = -2(c_{11} + 3 c_\delta c_{23})^3$. Clearly then $f=0$,
and the conditions for degeneracy at leading order are
\be
c_{12} = c_{13} \,,\qquad c_{12}^2 = c_{23}^2 + c_\delta c_{11}c_{23} \,.
\ee
Thus there is a two-parameter set of degeneracies at leading order for
this texture; at leading order $h_{eff}$ has the form
\be
h_{eff} = \pmatrix{c_{11} & S & Sc_\delta \cr S & 0 & c_{23} \cr
Sc_\delta & c_{23} & 0}E \,,
\ee
where $S \equiv \sqrt{c_{23}(c_{23}+c_\delta c_{11})}$. By applying the
rotation
\be
V = \pmatrix{1 & 0 & 0 \cr 0 & {1\over\sqrt2} & {1\over\sqrt2}c_\delta \cr
0 & - {1\over\sqrt2}c_\delta & {1\over\sqrt2}} \,,
\ee
and adding a term $c_\delta c_{23} E$ times the identity, at leading
order the new Hamiltonian is
\be
h_{eff}^\prime = V^T h_{eff} V = \pmatrix{
c_{11} + c_\delta c_{23} & 0 & \sqrt2 S c_\delta \cr
0 & 0 & 0 \cr \sqrt2 S c_\delta & 0 & 2 c_\delta c_{23}}E \,.
\label{eq:4D}
\ee
Equation~(\ref{eq:4D}) has the form of Class 3B with $r = \sqrt{2c_{23}/
(c_{23} + c_\delta c_{11})}$. The matrix that diagonalizes the original
$h_{eff}$ is therefore $U^\prime = V U$, or
\be
U'=\frac{1}{\sqrt{2}}
\pmatrix{
\sqrt{2} \cos\phi
& \sqrt{2} \sin\phi \cos\theta
& - \sqrt{2} \sin\phi \sin\theta
\cr
c_\delta \sin \phi
& \sin\theta - c_\delta \cos\phi \cos\theta
& \cos\theta + c_\delta \cos\phi \sin\theta
\cr
\cos\phi
& - c_\delta \sin\theta - \cos\phi \cos\theta
& - c_\delta \cos\theta + \cos\phi \sin\theta
} \,,
\ee
where \(U\) is from Eq.~(\ref{eq:U3B}). The oscillation probabilities are
\bea
P(\nu_\mu \rightarrow \nu_\mu) &=& 1 -
\left(\sin\theta - \cos\phi c_\delta \cos\theta \right)^2
\left(\cos\theta + \cos\phi c_\delta \sin\theta \right)^2
\sin^2\left(\Delta_{23} \frac{L}{2}\right)
\nonumber \\
&&- \sin^2\phi \left(\sin\theta - c_\delta \cos\phi \cos\theta \right)^2
\sin^2\left(\Delta_{12} \frac{L}{2}\right)
\nonumber \\
&&- \sin^2\phi \left(\cos\theta - c_\delta \cos\phi \sin\theta \right)^2
\sin^2\left(\Delta_{13} \frac{L}{2}\right)\,,
\\
P(\nu_\mu \rightarrow \nu_e) &=&
\sin^2\phi \sin2\theta (\sin^2\phi \sin\theta \cos\theta
- c_\delta \cos\phi \cos2\theta)
\sin^2\left(\Delta_{23} \frac{L}{2}\right)
\nonumber \\
&&- c_\delta \sin\phi \sin2\phi (\sin\theta \cos\theta
- c_\delta \cos\phi \cos^2\theta) \sin^2\left(\Delta_{12} \frac{L}{2}\right)
 \nonumber \\
&&+ c_\delta \sin\phi \sin2\phi (\sin\theta \cos\theta
+ c_\delta \cos\phi \sin^2\theta) \sin^2\left(\Delta_{13} \frac{L}{2}\right)\,,
\\
P(\nu_e \rightarrow \nu_e) &=& 1 -
\sin^4\phi \sin^22\theta \sin^2\left(\Delta_{23} \frac{L}{2}\right)
- \sin^22\phi \cos^2\theta \sin^2\left(\Delta_{12} \frac{L}{2}\right)
\nonumber \\
&&- \sin^22\phi \sin^2\theta \sin^2\left(\Delta_{13} \frac{L}{2}\right)
\,.
\eea

In order to compare with the atmospheric and long-baseline neutrinos
data, for large E, we should have \(\Delta_{12}L,\Delta_{13}L\gg
\Delta_{23}L\sim 1\). Then the oscillation probabilities are
\bea
P(\nu_\mu \rightarrow \nu_\mu) &=& 1 -
\left(\sin \theta - c_\delta \cos\phi \cos\theta \right)^2
\left(\cos \theta + c_\delta \cos\phi \sin\theta \right)^2
\sin^2\left(\Delta_{23} \frac{L}{2}\right)
\nonumber \\
&&- \frac{1}{2}\sin^2 \phi (1+\cos^2 \phi)\,,
\\
P(\nu_\mu \rightarrow \nu_e)& =&\sin^2\phi \sin2\theta (\sin^2\phi \sin\theta \cos\theta
- c_\delta \cos\phi\cos2\theta)
\sin^2\left(\Delta_{23} \frac{L}{2}\right)+ \frac{1}{4} \sin^22\phi \,.
\eea
Maximal \(\nu_e\) oscillations requires
\bea
1 &\simeq&
\left(\sin\theta - c_\delta \cos\phi \cos\theta \right)^2
\left(\cos\theta + c_\delta \cos\phi \sin\theta \right)^2
\nonumber \\
&=& 1 - \sin^2\phi
- (\cos^2\phi + \frac{1}{2}c_\delta \cos\phi - \frac{1}{4} \sin^4\phi)
\sin^22\theta 
\eea

If \(\phi\) is small and \(\sin2\theta\simeq 0\), the probabilities
are appropriate for the atmospheric and long-baseline neutrinos. Since
\(\sin\phi \equiv r/\sqrt{1+r^2}\) and \(\tan\theta \equiv
\sqrt{1+r^2}|a_{12}|/|r a_{11}-a_{13}|\), this imposes the conditions:
(i) \(r \simeq 0\) and \(a_{12} \simeq 0\), or (ii) \(r \simeq 0\) and
\(r a_{11} \simeq a_{13} \).

\begin{figure}
\small
\centering
\includegraphics[width=10cm]{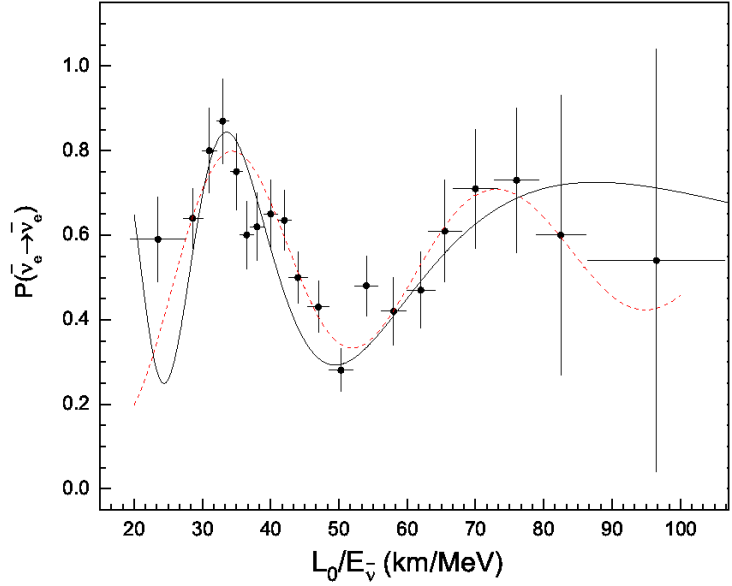}
\caption{Class 4D (solid lines) and the standard scenario (dashed lines)
compared to the KamLAND data. The model parameters are $a_{11} =
-9.0\times10^{-11}$~eV, $a_{12}=0$, $a_{13}=-9\times10^{-11}$~eV and \(r=0.02\).}
\label{fig:4DKamLAND}
\end{figure}

\begin{figure}
\small
\centering
\includegraphics[width=10cm]{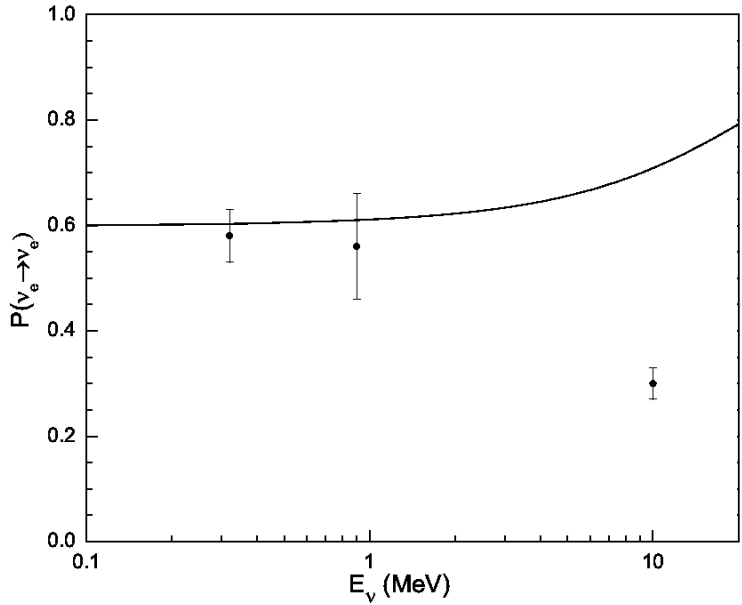}
\caption{The best fit prediction of Class 4D to the solar data. The model
parameters are $a_{11} = 9.0\times10^{-11}$~eV, $a_{12} = 0$,
$a_{13} = 9\times10^{-11}$~eV and $r = 0.02$.}
\label{fig:4Dsolar1}
\end{figure}

Since this case is equivalent to Class 3B after a rotation in the
\(\nu_\mu-\nu_\tau\) sector, the \(\nu_e \rightarrow \nu_e\)
oscillation probability expression is still the same. The results are also
similar to Class 3B. While there are parameter values that yield reasonable
agreement with the KamLAND data (see Fig.~\ref{fig:4DKamLAND}),
they did not agree with the solar data at high energies (see
Fig.~\ref{fig:4Dsolar1}).

\begin{figure}
\small
\centering
\includegraphics[width=10cm]{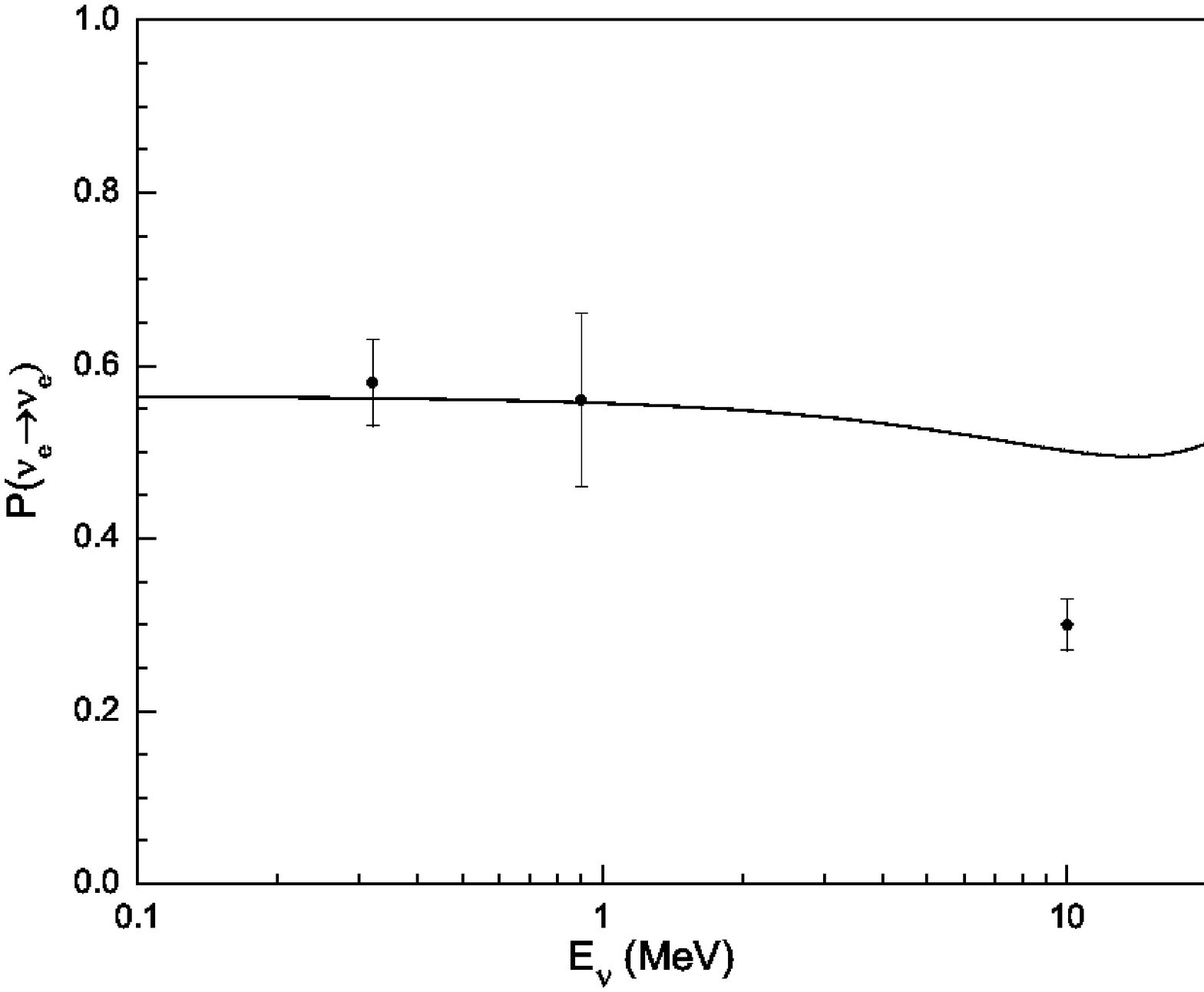}
\caption{Best fit prediction for survival probability of solar \(\nu_{e}\)
for Class 4D. The model parameters are
$a_{11} = -6.0\times10^{-11}$~eV, $a_{12} = 7.0\times10^{-11}$~eV,
$a_{13}=0.42\times10^{-11}$~eV and $r=-0.07$.}
\label{fig:4Dsolar2}
\end{figure}

Also, we fit the solar data separately. 
As was the case for Class 3B, we do not find a good fit
to the solar data at high energies (see Fig.~\ref{fig:4Dsolar2}).

\subsection{Five $c$ parameters}

\subsubsection{Class 5A}

This case has three diagonal and two off-diagonal nonzero $c$. By subtracting
a piece proportional to the identity, this case may be reduced to 4B or 4C.

\subsubsection{Class 5B}

This case has the structure
\bea
h_{eff} = \pmatrix{c_{11}E + a_{11} & c_{12}E + a_{12} & c_{13}E + a_{13} \cr
c_{12}^*E + a_{12}^* & a_{22} & c_{23}E + a_{23} \cr
c_{13}E + a_{13}^* & c_{23}E + a_{23}^* & c_{33}E + a_{33}} \,,
\eea
where $c_{11}$ and $c_{33}$ are real, $c_{13}$ and $c_{23}$ may be taken as
real and positive, and $c_{12}$ is complex. At leading order the cubic
equation for the eigenvalues of $h_{eff}/E$ is
\be
\lambda^3 - \lambda^2(c_{11} + c_{33}) + \lambda(c_{11}c_{33} - c_{12}^2
-c_{13}^2 - c_{23}^2) + c_{11}c_{23}^2 + c_{22}c_{13}^2
- 2c_{12}c_{13}c_{23}c_\delta \,,
\ee
where the $c_{12}$ is the magnitude and $\delta$ the phase of $c_{12}$.
Then we have
\bea
q &=& c_{11}^2 + c_{33}^2 - c_{11}c_{33} + 3(c_{12}^2 + c_{13}^2 + c_{23}^2)\,,
\\
r &=& -2(c_{11} + c_{33})^3 + 9c_{11}c_{33}(c_{11}+c_{33})
+ 9c_{11}(2 c_{23}^2 - c_{12}^2 - c_{13}^2)
\\
&& + 9c_{33}(2 c_{12}^2 - c_{13}^2 - c_{23}^2) -54 c_{12}c_{13}c_{23}c_\delta
\,.
\eea
and the conditions for a minimum of $f = 4q^3 - r^2$ are
\bea
0 &=& {\partial f\over\partial c_{11}} =
12 (2c_{11}-c_{33}) q^2 - 2r \left[-6c_{11}^2  + 6c_{11}c_{33} + 3 c_{33}^2
+ 9(2 c_{23}^2 - c_{12}^2 - c_{13}^2) \right] \,,
\label{eq:1c}\\
0 &=& {\partial f\over\partial c_{33}} =
12 (2c_{33}-c_{11}) q^2 - 2r \left[-6c_{33}^2  + 6c_{11}c_{33} + 3 c_{11}^2
+ 9(2 c_{12}^2 - c_{13}^2 - c_{23}^2) \right] \,,
\label{eq:2c}\\
0 &=& {\partial f\over\partial c_{13}} =
72 c_{13} q^2 + 36r \left[(c_{11}+c_{33})c_{13}
+ 3 c_{12}c_{23}c_\delta\right] \,,
\label{eq:3c}\\
0 &=& {\partial f\over\partial c_{12}} =
72 c_{12} q^2 + 36r \left[(c_{11}-2c_{33})c_{12}
+ 3 c_{13}c_{23}c_\delta\right] \,,
\label{eq:4c}\\
0 &=& {\partial f\over\partial c_{23}} =
72 c_{23} q^2 - 36r \left[(2c_{11}-c_{33})c_{23}
- 3 c_{12}c_{13}c_\delta \right] \,,
\label{eq:5c}\\
0 &=& {\partial f\over\partial\delta} =
-108 r c_{12}c_{13}c_{23}\sin\delta \,.
\label{eq:6c}
\eea

It can be shown by the usual arguments that $q$ and $r$ are not zero, and
$c_\delta = \pm1$. By eliminating $q$ and $r$ from
Eqs.~(\ref{eq:3c}) and (\ref{eq:4c}) we find
\be
c_{33}c_{12}c_{13} = c_\delta c_{23} (c_{13}^2 - c_{12}^2) \,,
\label{eq:7c}
\ee
and applying a simlar procedure to Eqs.~(\ref{eq:3c}) and (\ref{eq:5c}) gives
\be
c_{11}c_{13}c_{23} = c_\delta c_{12} (c_{13}^2 - c_{23}^2) \,.
\label{eq:8c}
\ee
Using the relations in Eqs.~(\ref{eq:7c}) and (\ref{eq:8c}) it can be shown
that all minimum conditions are met and that $f = 0$, so
Eqs.~(\ref{eq:7c}) and (\ref{eq:8c}) are the degeneracy conditions.
Thus there is a three-parameter set of degeneracies at leading order for this
texture. Then, after adding the term $c_{12}c_{23}c_\delta/c_{13}$ times the
identity, the effective Hamiltonian at leading order may be written as
\be
h_{eff} = {c_\delta\over c_{12}c_{13}c_{23}} \pmatrix{
c_{12}^2 c_{13}^2
& c_{13} c_{23} c_{12}^2
& c_\delta c_{13}^2 c_{12} c_{23}
\cr
c_{13} c_{23} c_{12}^2
& c_{12}^2 c_{23}^2
& c_\delta c_{12} c_{13} c_{23}^2 \cr
c_\delta c_{13}^2 c_{12} c_{23}
& c_\delta c_{12} c_{13} c_{23}^2
& c_{13}^2 c_{23}^2} E \,.
\ee
Without loss of generality $a_{22}$ may be set equal to zero. Then the
eigenvalues of $h_{eff}$ to order $E^0$ are
\be
\lambda_1 = {c_\delta S\over c_{12} c_{13} c_{23}}E + a_{11} + a_{33}
\,,\qquad
\lambda_2,\lambda_3 = {1\over2}\left[ x \pm \sqrt{x^2 - 4y} \right] \,,
\ee
where
\bea
x &\equiv& {1\over S}\left[ a_{11} c_{23}^2 (c_{12}^2 + c_{13}^2)
+ a_{33} c_{12}^2 (c_{13}^2 + c_{23}^2) \right.
\nonumber
\\
&\phantom{\equiv}& \phantom{{1\over S}} \left.
- 2{\rm Re}(a_{12}) c_{12}^2c_{13}c_{23}
- 2{\rm Re}(a_{13}) c_{13}^2c_{12}c_{23} c_\delta
- 2{\rm Re}(a_{23}) c_{12}c_{13}c_{23}^2 c_\delta \right]
 \,,
\\
y &\equiv& {1\over S^2} \left[
-2a_{11}{\rm Re}(a_{23})c_{12}c_{13}c_{23}^2 c_\delta
-2a_{33}{\rm Re}(a_{12})c_{13}c_{23}c_{12}^2 + a_{11}a_{33} c_{12}^2 c_{23}^2
\right.
\nonumber
\\
&\phantom{\equiv}& \phantom{{1\over S}}
-c_{13}^2 c_{23}^2 |a_{12}|^2 - c_{12}^2 c_{13}^2 |a_{23}|^2
- c_{12}^2 c_{23}^2 |a_{13}|^2 + 2{\rm Re}(a_{13}a_{23}) c_{13}c_{23} c_{12}^2
\nonumber
\\
&\phantom{\equiv}& \phantom{{1\over S}} \left.
+ 2{\rm Re}(a_{13}a_{12}^*) c_{12}c_{13} c_{23}^2 c_\delta
+ 2{\rm Re}(a_{23}a_{12}^*) c_{12}c_{23} c_{13}^2 c_\delta \right] \,,
\eea
and $S\equiv c_{12}^2 c_{13}^2 + c_{12}^2 c_{23}^2 + c_{13}^2 c_{23}^2$.
Thus degeneracy requires that the quadratic discriminant $g = x^2- 4y$ be zero.
It can be shown that $g$ has a minimum at zero when
\bea
y_{12} &=&
c_\delta c_{12}\left(\frac{y_{13}}{c_{13}}-\frac{y_{23}}{c_{23}}\right) \,,
\\
x_{12} &=& \frac{c_{23}a_{11}}{2c_{13}}
+ c_\delta \frac{c_{12}(c_{13}^2+c_{23}^2)}{c_{23}(c_{12}^2+c_{13}^2)}
\left(x_{23}-\frac{c_\delta c_{12} c_{33}} {2c_{13}}\right) \,,
\\
x_{13} &=& c_\delta \frac{ c_{12}^2 a_{33}+ c_{23}^2 a_{11}}{2c_{12}c_{23}}
+ \frac{c_{13}(c_{12}^2+c_{23}^2)}{c_{23}(c_{12}^2+c_{13}^2)}
\left(x_{23}-\frac{c_\delta c_{12} c_{33}} {2c_{13}}\right) \,,
\label{eq:gmin}
\eea
where \(x_{ij} = {\rm Re}(a_{ij})\), and \(y_{ij} = {\rm Im}(a_{ij})\);
these are the degeneracy conditions. The eigenvalues to order $E^{-1}$ are then
\bea
\lambda_1 &=& {c_\delta S E \over c_{12} c_{13} c_{23}} +
- {1\over SE}c_\delta c_{12} c_{13} c_{23}
\left[ a_{11} a_{33} - |a_{12}|^2 - |a_{13}|^2 - |a_{23}|^2\right] \,,
\\
\lambda_2 &=& {1\over SE}c_\delta c_{12} c_{13} c_{23}
\left[ a_{11} a_{33} - |a_{12}|^2 - |a_{13}|^2 - |a_{23}|^2\right] \,, \qquad
\lambda_3 = 0 \,,
\eea
and to leading order the mixing matrix that diagonalizes $h_{eff}$ is
\be
U = \pmatrix{
\cos\theta\cos\phi
& - \cos\xi\sin\phi - \sin\xi\sin\theta\cos\phi
& \sin\xi\sin\phi - \cos\xi\sin\theta\cos\phi
\cr
\sin\theta
& \cos\theta\sin\xi
& \cos\theta\cos\xi
\cr
\cos\theta\sin\phi
& \cos\xi\cos\phi - \sin\xi\sin\theta\sin\phi
&
-\sin\xi\cos\phi - \cos\xi\sin\theta\sin\phi
} \,,
\ee
where
\bea
\sin\theta &\equiv& {1\over\sqrt{S}} c_{12} c_{23} \,,\qquad
\sin\phi \equiv {c_{23} \over \sqrt{c_{12}^2 + c_{23}^2}} \,,
\\
\sin\xi &\equiv& c_\delta {c_{12}^2 c_{23}^2 (a_{11} + a_{33}) \over
N_3 \sqrt{c_{12}^2 + c_{23}^2}} \,,\qquad
\cos\xi = {\sqrt{S}\over N_3} {c_{23}^2 a_{11} - c_{12}^2 a_{33} \over
\sqrt{c_{23}^2 + c_{12}^2}} \,,
\label{eq:xi}
\eea
and
\be
N_3^2 = a_{33}^2 c_{12}^4 (c_{13}^2 + c_{23}^2) +
a_{11}^2 c_{23}^4 (c_{12}^2 + c_{13}^2) - 2 a_{11} a_{33} c_{13}^2
c_{12}^2 c_{23}^2 \,,
\ee
is a normalization factor. The oscillation probabilities are
\bea
P(\nu_\mu \to \nu_\mu) &=& 1
- \sin^2\xi\sin^22\theta \sin^2\left(\Delta_{21} {L\over 2} \right)
- \cos^2\xi\sin^22\theta \sin^2\left(\Delta_{31} {L\over 2} \right)
\nonumber
\\
&\phantom{=}& \phantom{1}
-\cos^4\theta \sin^22\xi \sin^2 \left(\Delta_{32} {L\over2} \right) \,,
\\
P(\nu_\mu \to \nu_e) &=&
2\sin2\theta\cos\theta \cos\phi\sin\xi
(\cos\xi\sin\phi+\sin\xi\sin\theta\cos\phi)
\sin^2\left(\Delta_{21} {L\over 2} \right)
\nonumber
\\
&\phantom{=}&
-2\sin2\theta\cos\theta \cos\phi\cos\xi
(\sin\xi\sin\phi-\cos\xi\sin\theta\cos\phi)
\sin^2\left(\Delta_{31} {L\over 2} \right)
\nonumber
\\
&\phantom{=}&
+2\sin2\xi\cos^2\theta (\cos\xi\sin\phi+\sin\xi\sin\theta\cos\phi)
(\sin\xi\sin\phi-\cos\xi\sin\theta\cos\phi)
\nonumber
\\
&\phantom{=}& \times \sin^2\left(\Delta_{32} {L\over 2} \right) \,.
\eea

In order to have nearly maximal $\nu_\mu$ oscillations at the
atmospheric scale, the $\Delta_{32}$ term must have amplitude
close to unity, or $\theta \simeq 0, \pi$ and $\xi \simeq \pi/4$.
Then from Eq.~(\ref{eq:xi})
\be
a_{33} c_{12} (c_{12} + c_\delta c_{23} \sin\theta) \simeq
a_{11} c_{23} (c_{23} - c_\delta c_{12} \sin\theta) \,.
\label{eq:xi2}
\ee
The small value of $\sin\theta$ implies $c_{12}^2, c_{23}^2 \ll c_{13}^2$.
Furthermore, in order to have small $\nu_\mu \to \nu_e$ oscillations
at the $\Delta_{32}$ scale, $\sin^2\phi \ll 1$, or $c_{23}^2 \ll
c_{12}^2$, {\it i.e.}, there is a hierarchy among the off-diagonal $c_{ij}$.
Then Eq.~(\ref{eq:xi2}) implies $a_{33} \ll a_{11}$ as well. Therefore
there is a lot of fine tuning required to achieve the proper mixing.

For simplicity, we only considered the parameters as real numbers.  We
have scanned the \(c_{12}\), \(c_{13}\) and \(c_{23}\) parameter
space to fit the KamLAND and solar data. Other parameters in the
Hamiltonian can be determined by these three parameters, {\it i.e.},
\(c_{11}\) and \(c_{13}\) can be determined from Eqs.~(\ref{eq:7c}) and
(\ref{eq:8c}). Also, for the atmospheric and long-baseline neutrinos, the
\(\Delta_{23}\) term has the correct energy dependence, and gives
\begin{equation}
\delta m_{eff}^2 = 2E\Delta_{23} =
\frac{2}{S} c_\delta c_{12}c_{13}c_{23}
[a_{11}a_{33}-|a_{12}|^2-|a_{13}|^2-|a_{23}|^2] \,.
\end{equation}
The above equation together with Eqs.~(\ref{eq:gmin}) and (\ref{eq:xi2})
 determine all \(a_{ij}\). Another constraint is the hierarchy among
the off-diagonal \(c_{ij}\), \(c_{23}^2 \ll c_{12}^2 \ll c_{13}^2\),
which is also considered during the parameter search.

\begin{figure}
\small
\centering
\includegraphics[width=10cm]{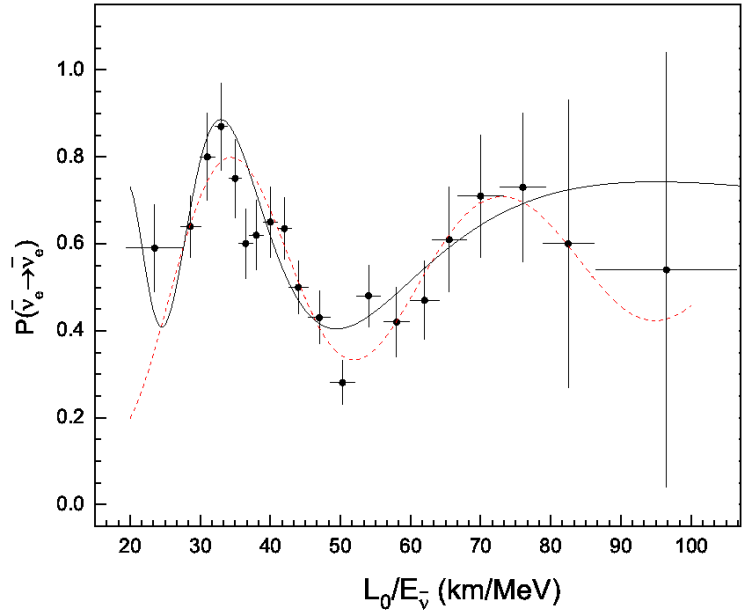}
\caption{Best fits for Class 5B (solid lines) and for standard oscillations (dashed
lines) compared to the KamLAND data. The model parameters are $c_{23} =
5.9\times10^{-23}$, $c_{12} = 1.0\times10^{-22}$ and $c_{13} =
8.9\times10^{-19}$. }
\label{fig:5BKamLAND}
\end{figure}

\begin{figure}
\small
\centering
\includegraphics[width=10cm]{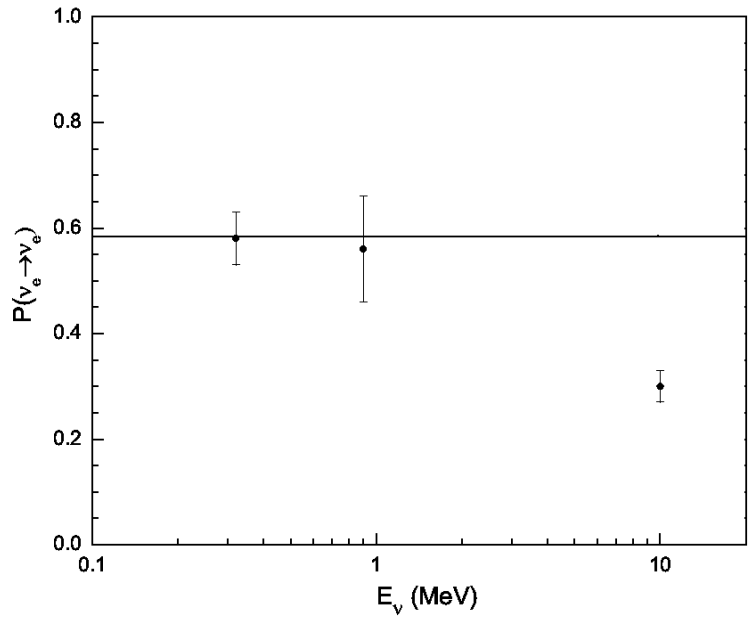}
\caption{The prediction of Class 5B for the solar neutrino survival probability
using the parameter values obtained from fitting the KamLAND data. }
\label{fig:5Bsolar1}
\end{figure}

\begin{figure}[!htb]
\small
\centering
\includegraphics[width=10cm]{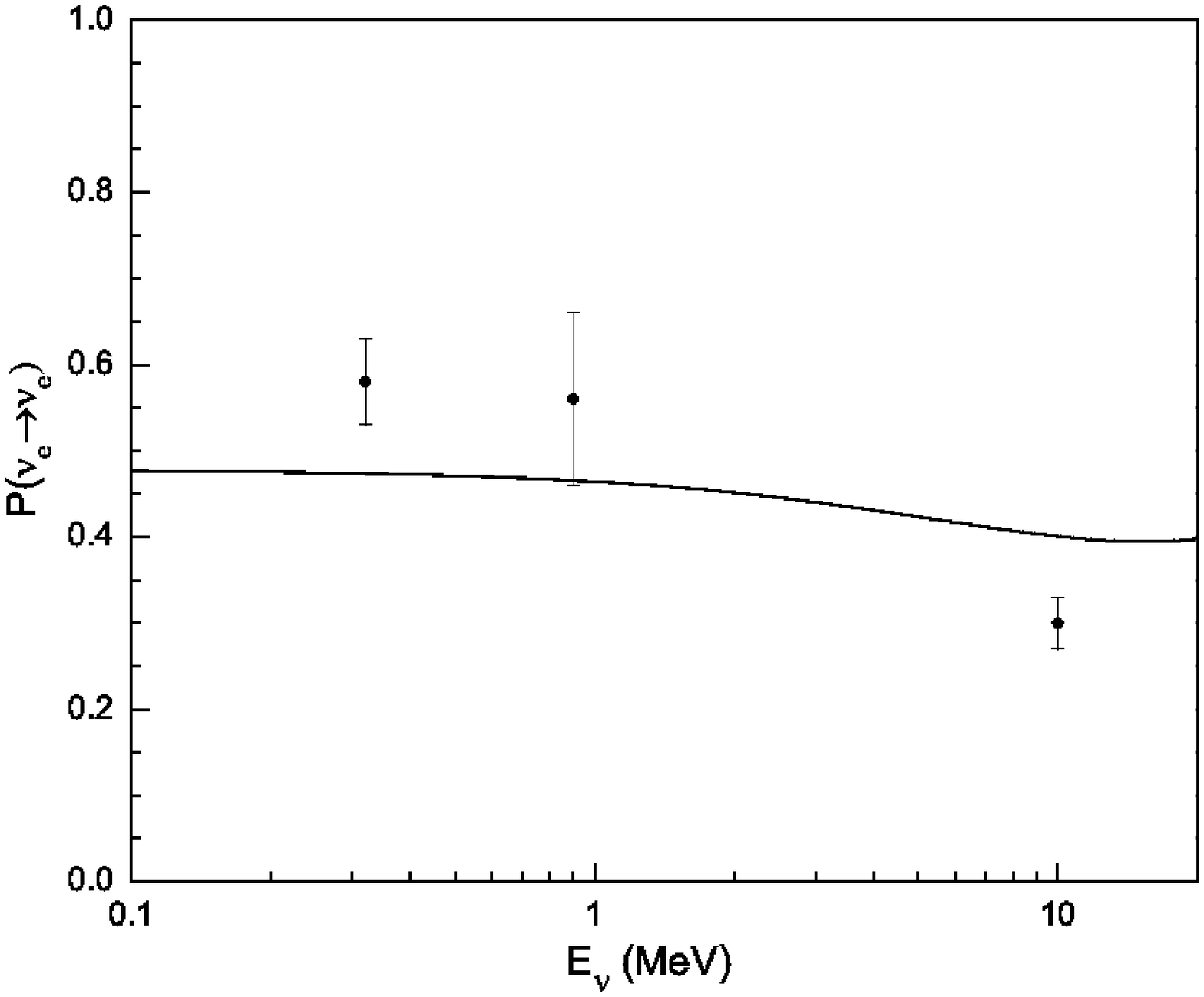}
\caption{Best fit prediction for the \(\nu_{e}\) survival probability in Class
5B. The model parameters for the best fit are $c_{23} = 1.0\times10^{-18}$,
$c_{12} = 2.4\times10^{-18}$ and $c_{13} = 1.0\times10^{-18}$.}
\label{fig:5Bsolar2}
\end{figure}

We have varied the range of \(c_{13}\) from the order of \(10^{-20}\)
to \(10^{-16}\) and take \(c_{12}\) and \(c_{23}\) to be at least one
order of magnitude less than \(c_{13}\) and \(c_{12}\)
respectively. We found parameter values that can fit the KamLAND data
(see Fig.~\ref{fig:5BKamLAND}), but they do not yield reasonable
agreement with the solar data at high energies (see
Fig.~\ref{fig:5Bsolar1}). We also attempted to fit solar neutrinos
alone and found there are no parameter values that can yield
reasonable agreement with the solar data. The best fit is shown in
Fig.~\ref{fig:5Bsolar2}.

\subsection{Six $c$ parameters}

In this case all $c$ elements are nonzero. By subtracting off a
piece proportional to the identity, this case may be reduced to Class
5B, which is ruled out.

\section{Summary}

We have examined the general three neutrino effective Hamiltonian in
Eq.~(\ref{eq:heff}) for the case of direction-independent interactions
and no neutrino mass. We looked for texture classes in which two eigenvalues
were degenerate to order $1/E$ at high neutrino energy, so that
oscillations of atmospheric and long-baseline neutrinos would exhibit
the usual $L/E$ dependence.

Among the classes that had the proper $1/E$ dependence at high energy,
none was also able to fit the atmospheric, long-baseline, solar and
KamLAND data simultaneously. Class 1A (along with the equivalent Classes
2A and 3A) reduced to the direction-independent bicycle model, which has
been shown to be inconsistent with the solar, atmospheric and long-baseline
neutrino data. Classes 2C (and the equivalent 3E) and 3F did not have the
proper oscillation amplitudes for atmospheric neutrinos. Finally, Classes
3B (and the equivalent Classes 3C, 4A and 4D) and 5B (and the equivalent
Class 6) were able to fit atmospheric and long-baseline neutrino data, but
could not simultaneously fit KamLAND and solar data at lower neutrino
energies. The major difficulty in these latter classes was reproducing the
low survival probability of high-energy solar neutrinos.

Although we have not made an exhaustive search of the parameter space,
the fact that high-energy neutrinos exhibit an $L/E$ dependence in
their oscillations over many orders of magnitude in $E$ suggests that
the only way this can occur in the effective Hamiltonian described by
Eq.~(\ref{eq:heff}) is via the degeneracy of two eigenvalues to order
$1/E$. Since none of the cases where such a degeneracy occurs are also
able to fit all neutrino data simultaneously, it seems extremely
unlikely that any direction-independent SME model
without neutrino mass will provide a viable description of all
neutrino oscillation phenmomena. There is also strong evidence against
direction-dependent terms.  Furthermore, nonrenormalizable Lorentz
noninvariant effective Hamiltonians with higher powers of energy (as
in, {\it e.g.}, the model of Ref.~\cite{Diaz}) and no neutrino masses
would require additional degeneracy conditions. Therefore it appears
highly unlikely that Lorentz invariance violation alone can account for
all of the observed oscillation phenomena.

\section*{Acknowledgments}

We thank Wan-yu Ye for computational assistance in the early stages of
this work and A. Kostelecky for useful discussions. We also
thank the Aspen Center for Physics for its hospitality during the initial
stages of this work. This research was supported by the
U.S. Department of Energy under Grant Nos.~DE-FG02-95ER40896,
DE-FG02-01ER41155, and DE-FG02-04ER41308, by the NSF under Grant 
No. PHY-0544278, and by the Wisconsin Alumni Research
Foundation.

\end{document}